\documentclass[11pt,a4paper]{article}
\usepackage[numbers]{natbib}
\usepackage{amsmath}
\usepackage{amssymb}
\usepackage{amsthm}
\usepackage[margin=2.1cm]{geometry}
\usepackage{color}
\usepackage{microtype}
\usepackage{graphicx}
\usepackage{standalone}
\usepackage{tikz}
\usepackage{enumerate}
\usepackage{mathrsfs}
\usepackage{BOONDOX-ds} 
\usepackage{tensor}
\usepackage{subcaption}
\usepackage[
singlelinecheck=false 
]{caption}
\usepackage{scalerel}
\usepackage{mathtools}

\newcommand*{\paral}{{\stretchrel*{\parallel}{\perp}}}
\newcommand{\delparal}{\nabla_{\!\paral}}
\newcommand{\dxperp}{\partial_{x_{\!\perp}}}

\newcommand{\RR}{\mathbb R}
\newcommand{\fracD}[3][0]{\tensor*[^{}_{#1}]{D}{^{#3}_{#2}}}
\newcommand{\fracDtilde}[3][0]{\tensor*[^{}_{#1}]{\tilde D}{^{#3}_{#2}}}
\graphicspath{{./}}

\newcommand{\Co}{\mathrm{Co}}

\newtheorem{thm}{Theorem}
\newtheorem{defi}{Definition}
\newtheorem{lem}{Lemma}

\theoremstyle{remark}
\newtheorem{rmrk}{Remark}
\usepackage{color}
\definecolor{comment}{RGB}{65,104,23}

\newcommand{\at}{\makeatletter @\makeatother}

\makeatletter
\renewcommand{\maketitle}{\bgroup\setlength{\parindent}{0pt}
\begin{flushleft}
  \textbf{\Large\@title}\par
	\vspace{.3cm}
  \@author\par
\noindent\rule[0.5ex]{\linewidth}{0.5pt}
\end{flushleft}\egroup
}
\makeatother

\title{Investigations of an effective time-domain boundary condition for quiescent  viscothermal acoustics}
\author{\underline{{LINUS HÄGG} and MARTIN BERGGREN}\\
Department of Computing Science,
Umeå University, SE--901 87 Umeå, Sweden.\\
\{linush, martin.berggren\}\raisebox{-1pt}{{\at}}cs.umu.se\\
\today}

\begin{document}
\thispagestyle{empty}
\maketitle

\begin{abstract} 
Accurate simulations of sound propagation in narrow geometries need to account for viscous and thermal losses.
In this respect, effective boundary conditions that model viscothermal losses in frequency-domain acoustics have recently gained in popularity.
Here, we investigate the time-domain analogue of one such boundary condition. 
We find that the thermal part of the boundary condition is passive in time domain as expected, while the viscous part is not. 
More precisely, we demonstrate that the viscous part is responsible for exponentially growing normal modes with unbounded temporal growth rates, which indicates ill-posedness of the considered model.
A finite-difference-time-domain scheme is developed for simulations of lossy sound propagation in a duct. 
If viscous losses are neglected the obtained transmission characteristics are found to be in excellent agreement with frequency-domain simulations. 
In the general case, the simulations experience an instability much in line with the theoretical findings.  

\end{abstract}
\section{Introduction}
For acoustic wave propagation in fluids, viscous and thermal losses are typically small and concentrated in the so-called viscous and thermal boundary layers close to solid walls.
The thicknesses of the boundary layers depend on properties of the medium in which the sound propagates and the frequency of the sound. 
For sound propagation in air, the thicknesses of the viscous and thermal boundary layers are of the same order, ranging from about $15$ {\textmu}m at $20$ kHz to $0.5$ mm at $20$ Hz. 
Although often otherwise negligible, the effects of viscous boundary losses may be significant in narrow geometries.

The frequent use of miniaturized acoustic devices, such as portable headphones, mobile phones, and hearing aids, has increased the need for accurate and computationally efficient models of viscothermal acoustics.  
Assuming that nonlinear effects are negligible, the linearized, compressible Navier--Stokes equations are the fundamental model for viscothermal (also known as \textit{thermoviscous}) acoustics in general.
However, the extreme thinness of the viscothermal boundary layer in air, compared to the free space wavelength and characteristic dimensions of relevant geometries, implies that it can be computationally expensive to generate accurate numerical solutions to the linearized, compressible Navier--Stokes equations.
This can be understood from that the above mentioned boundary-layer thicknesses are smaller than the wavelength by a factor of approximately $10^{-6}$--$10^{-3}$ (low to high frequency).

As is often true in applications, we assume that the mean flow is slow compared to the speed of sound, so that it is justified to linearize the compressible Navier--Stokes equations around a fluid state at rest, which leads to a considerable simplification of the resulting equations.  
Since Kirchhoff~\cite{Ki68} devised a semi-analytical solution for lossy sound propagation in a cylindrical pipe, much effort has been devoted to reformulations, approximate models, or numerical schemes that may reduce the computational cost of viscothermal acoustic simulations. 
Chabassier and Thibault~\cite{ChTh20} provide a recent comprehensive review of existing approaches with emphasis on simulations of wind instruments in frequency domain.
Some approaches~\cite{Jo10, KaWiBo11} improve the computational efficiency by reducing the number of equations; nonetheless, the need to resolve the boundary layers persists. 
Other approaches, based on, for instance, the Zwikker--Kosten or Webster--Lokshin models, have been developed for special geometries in which sound propagation is assumed to be (approximately) one-dimensional~\cite{ChTh20}. 
A third approach, based on the observation that viscothermal losses are typically small and concentrated \emph{close} to the walls of the domain, is to artificially concentrate the losses \emph{at} the walls using an effective boundary condition, which may be derived using acoustic boundary layer theory. 
In general, such effective boundary conditions are applicable when the boundary layer thicknesses are small compared to the free space wavelength and characteristic dimension of the geometry~\cite{ChTh20}.
Effective boundary conditions have proven to be both accurate and computationally efficient in frequency-domain simulations involving complex geometries~\cite{BaBr18, BeBeNo18, JiSa18}, even for the geometries of microperforated plates and fibrous materials, which at places are so narrow that the underlying model is barely applicable~\cite{CoDaMaBaBe20, BiTiGaVe21}. 
Recently, effective boundary conditions of this type have been implemented in the Acoustics Module of the commercial software COMSOL Multiphysics.        

The idea of using boundary layer theory to devise approximate acoustic models that are appropriate close to the walls goes back to Cremer~\cite{Cr48}, who derived the dimensionless wall admittance $\Upsilon_w$ of a plane wave ($\exp(i\omega t - ik\cdot x)$, $\omega = c|k|>0$) impinging at an angle $\theta$ on an infinite planar wall with exterior unit normal $n$,
\begin{equation}
	\label{cremer}
	\Upsilon_w = \sin^2\!\theta\,\sqrt{i\omega\tau_V} + \sqrt{i\omega\tau_T} = \frac{|k|^2-(n\cdot k)^2}{|k|^2}\sqrt{i\omega\tau_V} + \sqrt{i\omega\tau_T},
\end{equation}
where $\sqrt{i} = (1+i)/\sqrt{2}$. 
We have chosen here to express Cremer's formula for the admittance using viscous and thermal time scales,
\begin{align}
	\label{tauVAndT}
	\tau_V = \frac{\nu}{c^2}\text{ and }\tau_T = \frac{(\gamma-1)^2\kappa}{\rho_0 c^2\,c_p},
\end{align}
where $\rho_0$ denotes the ambient mass density, $c$ the speed of sound, $\nu$ the kinematic viscosity, $\kappa$ the thermal conductivity, $\gamma$ the heat capacity ratio, and $c_p$ the specific heat capacity at constant pressure.
\begin{rmrk}
More common is to specify boundary-layer models using the viscous and thermal boundary-layer \textit{thicknesses}
\begin{equation}
\delta_V = \sqrt{\frac{2\nu}{\omega}},
\qquad
\delta_T =\sqrt{\frac{2\kappa}{\omega\rho_0 c_p}}.
\end{equation}
However, formulations using time scales are useful for our final aim of devising a time-domain boundary-layer formulation.
\end{rmrk}
Table~\ref{t:matProp} in Appendix~\ref{A:matProp} presents properties of air employed in the numerical experiments that we report below. 
Comparing the timescales  $\tau_V \sim 10^{-10}$~s and $\tau_T \sim 10^{-11}$~s at atmospheric conditions to the frequencies of audible sound $f \in [20, 20\,000]$ Hz, we find that typically $\Upsilon_w\approx 0$, which approximates a slip condition on the acoustic velocity.
Pierce~\cite[eq.~(10-4.12)]{Pi89}, derived a generalization of expression~\eqref{cremer} in the form of a boundary condition for time-harmonic fields, which may be expressed using parameters~\eqref{tauVAndT} as
\begin{equation}
	\label{harmBLBC}
	\rho_0 c\,  n\cdot \hat u = -\frac{\rho_0 c^2\tau_V}{\sqrt{i\omega\tau_V}}\nabla_\Gamma\cdot \hat u_\Gamma + \sqrt{i\omega\tau_T}\hat p = -\frac{\sqrt{i\omega\tau_V}}{\omega^2}c^2\Delta_\Gamma \hat p + \sqrt{i\omega\tau_T}\hat p,
\end{equation} 
where $\hat p$ denotes the complex acoustic pressure, $\Delta_\Gamma$ the tangential Laplacian, $\hat u$ the acoustic velocity satisfying $i\omega\rho_0\hat u = -\nabla \hat p$, and $\nabla_\Gamma\cdot \hat u_\Gamma$ the tangential divergence of the acoustic velocity.
Note that boundary condition~\eqref{harmBLBC} yields the dimensionless wall admittance~\eqref{cremer} for $\hat p(x) \propto \exp(-ik\cdot x)$ with $|k| = \omega/c$.

\begin{rmrk}
The tangential gradient $\nabla_\Gamma f$ of a scalar or vector field $f$ along a smooth orientable surface $\Gamma$ embedded in $\RR^3$ is the projection of the gradient in the tangential direction, that is,  $\nabla_\Gamma f = P_\Gamma\nabla f$, where $P_\Gamma = I - n\otimes n$ and where $n$ is a unit normal field on $\Gamma$. 
The tangential divergence $\nabla_\Gamma\cdot u$ of a vector field is $u$ is the trace of the second-order tensor $\nabla_\Gamma u$.
The tangential gradient and divergence also satisfy 
\begin{equation}
\begin{aligned}
\nabla_\Gamma f &= \nabla f - n\frac{\partial f}{\partial n}, 
\\
\nabla_\Gamma\cdot u &= \nabla\cdot u - n\cdot \frac{\partial u}{\partial n}.
\end{aligned}
\end{equation}  
The tangential Laplacian $\Delta_\Gamma$ is the tangential divergence of the tangential gradient.
\end{rmrk}

Over the years, the effective boundary conditions given by Cremer and Pierce have been independently rediscovered multiple times.   
Bossart et al.~\cite{BoJoBr03}, who appears to have been the first to exploit the potential for computational acoustics, propose a two step method coupling the dimensionless wall admittance~\eqref{cremer} with a lossy Helmholtz equation (that is, with a complex wave number) for the acoustic pressure in the bulk region. 
In the first step, the approximation $(|k|^2-(n\cdot k)^2)/|k|^{2} \approx 1/2$ leads to an approximation of the pressure $\hat p_1$, which in turn yields the approximation $(|k|^2-(n\cdot k)^2)/|k|^{2} \approx -c^2\Delta_\Gamma \hat p_1/(\omega^2 \hat p_1)$ used in the second step.
Later, one-step approaches based on coupling Pierce's boundary condition~\eqref{harmBLBC} with the Helmholtz equation have been successfully implemented~\cite{Ni10, ChLiGr13, BaBr18, BeBeNo18, JiSa18} (note that some of the references only consider viscous damping).
Berggren et al.~\cite{BeBeNo18} were the first to prove well-posedness of a variational formulation for the coupling of the Helmholtz equation with the boundary condition~\eqref{harmBLBC}. 

In case there are only viscous losses, some generalizations have been put forward. 
Cheng et al.~\cite{ChLiGr13} investigate sound propagation in a thin rectangular capillary with one of the dimensions comparable to the boundary layer thicknesses and devise a nonlocal effective boundary condition for this case.
The effects of curvature and other higher order effects of viscous losses have been investigated in the works of Schmidt et al.~\cite{ScThJo14,ScTh21}. 

So far, only frequency-domain approaches have been discussed.
Time-domain simulations may be computationally advantageous for determining broadband characteristics, especially if explicit time-stepping is possible.   
Moreover, a time-domain acoustic boundary condition has the advantage that it may be coupled to nonlinear equations such as the compressible Euler equations.
Time-domain approaches have been developed for both the Zwikker--Kosten and Webster--Lokshin models~\cite{ChTh20}.  
The aim of this article is to investigate the time-domain equivalent of the more general boundary condition~\eqref{harmBLBC}.
As will be described below, the presence of radicals will manifest itself in nonlocal, temporal, half-order integrals and derivatives, which require specialized discretizations in order to achieve sufficient computational efficiency~\cite{MoMaPiPa16}. 
The presence of tangential derivatives implies that boundary condition~\eqref{harmBLBC} is not locally reacting, and therefore not considered within the extensive framework of time-domain impedance boundary conditions~\cite{RiHaPaScBuTh11, Mo18}.
In isolation, the thermal part of boundary condition~\eqref{harmBLBC} can be analyzed within in that framework~\cite[§\,4.6]{Mo18}. 
However, the main advantage of using boundary condition~\eqref{harmBLBC}, compared to a generic time-domain impedance boundary condition, is that there are no free parameters that need tuning.  
           
As mentioned above, the aim of this article is to investigate the time-domain equivalent of boundary condition~\eqref{harmBLBC}. 
In Section~\ref{sec:TDVTBC}, we present the time-domain boundary condition and demonstrate that the viscous contribution is \emph{not} passive, contrary to the thermal contribution. 
In Theorem~\ref{th:1}, we reach the conclusion that the viscous part of boundary condition~\eqref{harmBLBC} allows solutions that grow arbitrary fast in time, which is a strong indication of ill posedness. 
This finding is somewhat surprising, given that the frequency-domain formulation has been proven to provide well-posed variational formulations~\cite{BeBeNo18, ScTh21}.
More precisely, the Briggs--Bers-type normal-mode analysis carried out in Section~\ref{sec:illposedness} indicates an absolute instability that becomes worse for increasing wave number.
This instability is likely the cause of the observed stationary grid-level oscillations associated with the highest representable wave number in the finite-difference implementation described in Sections~\ref{sec:FDTD} and~\ref{sec:NumExp}.
In order to obtain a stable time-domain model for the viscous boundary condition, either a stabilization scheme or an alternative model has yet to be worked out.
In contrast, when viscous losses are neglected, the simulated transmission characteristics of the planar mode is found to be in excellent agreement with frequency-domain simulations based on boundary condition~\eqref{harmBLBC}.        

\section{A time-domain viscothermal boundary condition}
\label{sec:TDVTBC}
Although the time-domain analogue of boundary condition~\eqref{harmBLBC} may be derived from scratch by mimicking, in time domain, the frequency-domain procedure outlined by Berggren et al.~\cite{BeBeNo18}, Fourier transforms~\cite[expressions~(7.1) and~(7.4)]{SaKiMa93} provide a more direct approach, 
\begin{equation}
	\label{BLBC}
	n\cdot u = - c\,\nabla_\Gamma\cdot\big(\sqrt{\tau_V} \fracD[-\infty]{t}{-1/2}u_\Gamma\big) +  \sqrt{\tau_T}\fracD[-\infty]{t}{1/2}  \frac{p}{\rho_0c}.
\end{equation}
Here, $\fracD[-\infty]{t}{-1/2}$ and $\fracD[-\infty]{t}{1/2}$ denote (left) temporal half-order Riemann--Liouville  fractional  integral and differential operators starting at $-\infty$ defined by
\begin{align}
	\label{defID}
	\fracD[-\infty]{t}{-1/2}p =  \frac{1}{\sqrt{\pi}}\int\limits_{-\infty}^{t}\frac{p(\tau)}{\sqrt{t - \tau}}\,\mathrm{d}\tau\enspace\text{ and }\enspace \fracD[-\infty]{t}{1/2}p = \partial_t \,\fracD[-\infty]{t}{-1/2}p,
\end{align}
respectively~\cite[§5.1~eq.~(5.2) and (5.6)]{SaKiMa93}.

Analogously as in frequency domain~\cite{BeBeNo18}, we propose to use boundary condition~\eqref{BLBC} in conjunction with the isentropic equations of sound propagation,
\begin{align}
	\label{isen}
	\partial_t p + \nabla\cdot (c\,u_p) &= 0,\\
	\partial_t u_p + c\nabla p &= 0,
\end{align}
where $u_p = \rho_0 c \,u$.

\subsection{Time-domain passivity}
\label{sec:TDpass}
\begin{figure}
	\centering
	\includegraphics{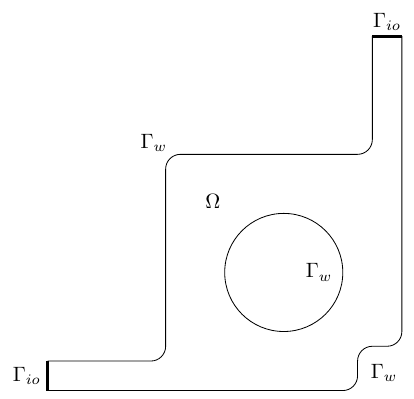}
	\caption{Illustration of the type of domain considered in initial--boundary-value problem~\eqref{ibp}.}
	\label{domain}
\end{figure}

In this section, we investigate the time-domain passivity of boundary condition~\eqref{BLBC}. 
To that end, we consider the initial--boundary-value problem
\begin{subequations}
	\label{ibp}
	\begin{align}
		\partial_t p + \nabla\cdot(cu_p) &= 0 &&\text{in }Q = (0,T)\times\Omega,\label{ibp1}\\
		\partial_t u_p + c\nabla p &= 0 && \text{in }Q = (0,T)\times\Omega,\label{ibp2}\\
		p = 0, u_p &= 0 && \text{on }Q_0 = \{0\}\times\Omega,\label{ibp3}\\
		n\cdot u_p  &=  -c\,\nabla_\Gamma\cdot\big(\sqrt{\tau_V}\, \fracD{t}{-1/2}u_{p,\Gamma}\big) +  \sqrt{\tau_T}\,\fracD{t}{1/2}  p && \text{on }\Sigma_w = (0,T)\times\Gamma_w, \label{ibp4}\\
		p - n\cdot u_p &= 2g &&  \text{on }\Sigma_{io} = (0,T)\times\Gamma_{io}\label{ibp5},
	\end{align}
\end{subequations}
where  $u_p = \rho_0 c\,u$ as before, $T>0$ is an arbitrary end time, and $g$ is a finite duration source acting at the in/out-boundary part $\Gamma_\text{io}$, which is complementary to the solid wall $\Gamma_w$ as illustrated in Figure~\ref{domain}.
Note that, since $p$ and $u_p$ vanish at $t = 0$ and $g$ is a finite duration source supported in $(0,T)$, we may assume that $p$ and $u_p$ vanish also for $t < 0$, which implies that $\fracD[-\infty]{t}{-1/2}u_{p,\Gamma} = \fracD{t}{-1/2}u_{p,\Gamma}$ and $\fracD[-\infty]{t}{1/2}p = \fracD[0]{t}{1/2} p$.
Assuming sufficient regularity of the domain and the acoustic fields, we apply $\int_\Omega p$ to equation~\eqref{ibp1}, $\int_\Omega u_p\cdot$ to equation~\eqref{ibp2}, sum the resulting terms, and integrate by parts either of the spatial derivatives, 
\begin{align}
	0 = \int\limits_\Omega \! p\big(\partial_t p + \nabla\cdot(cu_p) \big)\mathrm{d}\Omega + \int\limits_\Omega \! u_p\cdot\big(\partial_t u_p + c\nabla p\big)\mathrm{d}\Omega = \frac{1}{2}\frac{d}{dt}\int\limits_\Omega\! \big(p^2+|u_p|^2\big)\mathrm{d}\Omega + \int\limits_{\partial \Omega}\! cp\,n\cdot u_p\mathrm{d}\Gamma. 
\end{align}
Rearranging terms, integrating in time over $(0,T)$, invoking initial condition~\eqref{ibp3} and boundary condition~\eqref{ibp5}, we find
\begin{align}
	\frac{1}{2} \int\limits_\Omega \left(p^2 + |u_p|^2\right)\rvert_{t = T}\,\mathrm{d}\Omega &=  \int\limits_{\Sigma_{io}} cp(2g-p)\,\mathrm{d}\Sigma -\int\limits_{\Sigma_w}cp \, n\cdot u_p\,\mathrm{d}\Sigma\nonumber\\  
	&= \int\limits_{\Sigma_{io}} cg^2\,\mathrm{d}\Sigma -\int\limits_{\Sigma_{io}} c(p-g)^2\,\mathrm{d}\Sigma-\int\limits_{\Sigma_w}cp \, n\cdot u_p\,\mathrm{d}\Sigma
\end{align}
Thus, boundedness of the solution, in the form
\begin{equation}
	\label{energyEst}
	\int\limits_\Omega \left(p^2 + |u_p|^2\right)\rvert_{t = T}\,\mathrm{d}\Omega \leq \int\limits_{\Sigma_{io}} cg^2\,\mathrm{d}\Sigma
\end{equation}
would follow, provided that $\int_{\Sigma_w}cp \, n\cdot u_p\,\mathrm{d}\Sigma\geq 0$.
Strictly speaking, passivity requires that the bound~\eqref{energyEst} holds for each $T > 0$~\cite[Def.~3.3]{BeWo66}.
Invoking boundary condition~\eqref{ibp4}, we find that
\begin{equation}
	\int\limits_{\Sigma_w}cp \, n\cdot u_p\,\mathrm{d}\Sigma\ = -\int\limits_{\Sigma_w} c^2 p\,\nabla_\Gamma\cdot\big(\sqrt{\tau_V} \,\fracD{t}{-1/2}u_{p,\Gamma}\big) \,\mathrm{d}\Sigma +  \int\limits_{\Sigma_w}cp \,\sqrt{\tau_T}\,\fracD{t}{1/2}p\,\mathrm{d}\Sigma.
\end{equation}
Positivity of the quadratic form involving the half-derivative,  for any $T>0$, may be demonstrated using a diffusive representation~\cite[§2~\& Appendix~E]{Mo18} or, since the half-derivative may be expressed as a convolution whose Laplace transformed kernel is a positive-real function~\cite[eq.~(1.6)]{Mo18}, using systems theory~\cite{Ze63}.
Thus the thermal contribution is indeed passive. 

We proceed to investigate the viscous contribution. 
Inspired by the frequency-domain analysis~\cite{BeBeNo18}, we integrate by parts on the boundary to transfer the tangential derivatives to the pressure,
\begin{align}
	\label{exp1}
	-\int\limits_{\Sigma_w} c^2 p\,\nabla_\Gamma\cdot\big(\sqrt{\tau_V} \,\fracD{t}{-1/2}u_{p,\Gamma}\big) \,\mathrm{d}\Sigma &= \int\limits_{\Sigma_w} c^2 \nabla_\Gamma p\cdot \sqrt{\tau_V}\, \fracD{t}{-1/2}u_{p,\Gamma}\,\mathrm{d}\Sigma\nonumber\\ 
	 &\qquad- \int\limits_{\gamma_w} c^2 p\, \sqrt{\tau_V}\,\fracD{t}{-1/2} n_\Gamma\cdot u_{p,\Gamma}\,\mathrm{d}\gamma, 
\end{align}
where $\gamma_w = (0,T)\times \partial \Gamma_{w}$ and $n_\Gamma$ denotes the co-normal of $\Gamma_w$ at $\partial \Gamma_w$, that is, the  exterior unit normal on $\partial\Gamma_w$. 
Guided by the frequency-domain well-posedness theory~\cite{BeBeNo18}, we make the additional assumption that 
\begin{equation}
	\label{ass1}
	n_\Gamma\cdot u_{p,\Gamma} = 0 \qquad\text{on } \gamma_w = (0,T)\times \partial \Gamma_{w}.
\end{equation}
Then, using equation~\eqref{ibp2} to formally exchange the tangential gradient of the pressure with the tangential velocity $u_{p,\Gamma}$ in the first term on the right side of expression~\eqref{exp1}, we find
\begin{align}
	-\int\limits_{\Sigma_w} c^2 p\,\nabla_\Gamma\cdot\big(\sqrt{\tau_V} \,\fracD{t}{-1/2}u_p\big) \,\mathrm{d}\Sigma &= -\int\limits_{\Sigma_w} c \,\partial_t u_{p,\Gamma} \cdot \sqrt{\tau_V} \,\fracD{t}{-1/2}u_{p,\Gamma}\,\mathrm{d}\Sigma\nonumber\\ 
	&= -\int\limits_{\Sigma_w} c \,\partial_t u_{p,\Gamma} \cdot \sqrt{\tau_V} \,\fracD{t}{-3/2}\partial_t u_{p,\Gamma}\,\mathrm{d}\Sigma, 
\end{align}
where the last step follows since $\fracD{t}{-3/2} = \fracD{t}{-1/2}\fracD{t}{-1}$~\cite[Th.~2.5]{SaKiMa93} and $u_{p,\Gamma} = 0$ at $t = 0$ implies that $(\fracD{t}{-1} \partial_t u_{p,\Gamma})(t) =\int_{0}^t\partial_\tau u_{p,\Gamma}(\tau)\,\mathrm{d}\tau = u_{p,\Gamma}(t)-u_{p,\Gamma}(0) = u_{p,\Gamma}(t)$.
Thus, under assumption~\eqref{ass1}, the viscous part of boundary condition~\eqref{ibp4} is passive provided $\int_0^T \partial_t u_{p,\Gamma} \cdot\fracD{t}{-3/2}\partial_t u_{p,\Gamma}\,\mathrm{d}t$ is \emph{negative} for each $T>0$.
Recall the definition of the Riemann--Liouville fractional integral of order 3/2 starting at 0~\cite[§5.1 eq.~(5.1)]{SaKiMa93},   
\begin{equation}
	\fracD{t}{-3/2}v = \frac{2}{\sqrt \pi}\int \limits_0^t \sqrt{t-\tau} \,v(\tau)\,\mathrm{d}\tau.
\end{equation}
Note that $\int_0^T v(t) \fracD{t}{-3/2}v\,\mathrm{d}t > 0$ for non-vanishing functions $v$ that are either non-negative or non-positive.
In particular, there are such functions that are smooth and compactly supported in $(0,\infty)$, thus satisfying homogeneous initial conditions. 
Moreover, if $\int_0^T v(t) \fracD{t}{-3/2}v\,\mathrm{d}t < 0$ for some $v\in C([0,T])$ that vanishes at $t = 0$, then $v$ is non-vanishing and therefore either non-negative or non-positive in $(0,T^\ast)$ for some $T^\ast\in (0,T)$ and thus $\int_0^{T^\ast} v(t) \fracD{t}{-3/2}v\,\mathrm{d}t > 0$.   
Therefore negativity cannot hold for all $T>0$ and thus the viscous contribution to boundary condition~\eqref{ibp4} is not passive.       

To elucidate the issue, we invoke an expansion, proved in Theorem~\ref{th:expansion} in Appendix~\ref{sec:expansion},
\begin{align}
	\label{exp32}
	\beta_{3/2}\sqrt{\frac{2}{T}}\int\limits_0^T v(t) \fracD{t}{-3/2}v\,\mathrm{d}t = \left(\int\limits_0^T v(t)\,\mathrm{d}t \right)^2 - \sum\limits_{n = 1}^\infty (4n-1)\left(\int\limits_0^T P_n^{(-1/4)}\left(\frac{2t}{T}-1\right)v(t)\,\mathrm{d}t \right)^2, 
\end{align}
where $\beta_{3/2}>0$ and $P_n^{(-1/4)}$ denotes the $n$th order Gegenbauer polynomial with parameter $-1/4$.  
Although the expansion~\eqref{exp32} is \emph{not} orthogonal --- recall that the Gegenbauer polynomials are orthogonal with respect to a \emph{weighted} $L^2$ inner product --- we may conclude that $\int_0^T v(t) \fracD{t}{-3/2}v\,\mathrm{d}t \leq 0$ for functions averaging to zero on $(0,T)$.
Representing $u_{p,\Gamma} = (u_{p,\Gamma_1},\ldots,u_{p,\Gamma_{d}})$ in Cartesian coordinates, substituting $v = \partial_t u_{p,\Gamma_i}$ into the first term in expansion~\eqref{exp32}, and recalling initial condition~\eqref{ibp3}, we find
\begin{equation}
	\left(\int\limits_0^T \partial_t u_{p,\Gamma_i}\right)^2 =\big (u_{p,\Gamma_i}(T)-u_{p,\Gamma_i}(0)\big)^2 = u_{p,\Gamma_i}(T)^2, \qquad i\in\{1,\ldots,d\}.
\end{equation}
Thus, imposing the final condition $u_{p,\Gamma}(T) = 0$ on $\Sigma_w$ would let us conclude that the viscous contribution is \emph{passive on average} on $(0,T)$.  
We note that such final condition is fundamentally incompatible with an initial--boundary-value problem but compatible and implicitly incorporated in the time-harmonic setting. 

\subsection{Ill-posedness for  $\tau_V>0$}
\label{sec:illposedness}
In this section, we employ a Briggs--Bers-type analysis of normal modes~\cite{Br09} to demonstrate that supplementing system~\eqref{isen} with boundary condition~\eqref{BLBC} allows for normal modes that grow arbitrarily fast in time, provided that $\tau_V > 0$. 
As argued by Brambley~\cite{Br09}, unbounded exponential growth in time of normal modes is indicative of ill-posedness. 

\begin{figure}
	\centering
	\includegraphics{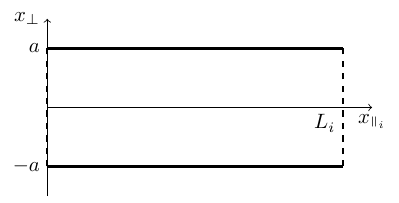}
	\caption{The existence of exponentially growing modes of arbitrary growth rates for $\tau_V>0$ is demonstrated for wave propagation between two parallel plates.}
	\label{f:parpl}
\end{figure}

We consider propagation in a right rectangular prism $\Omega_\paral \times (-a,a) = \Pi_{i=1}^{d-1}(0,L_i)\times (-a,a)$, which is bounded by two parallel plates located at $x_\perp = -a$ and $x_\perp = a$ as illustrated in Figure~\ref{f:parpl}.
Note that the coordinate system is aligned with the orientation of the plates so that $x = (x_\paral,x_\perp)\in\mathbb{R}^d$, where $x_\paral= (x_{\paral_1},\ldots,x_{\paral_{d-1}})$ is $(d-1)$-dimensional and tangential to the plates, and $x_\perp$ is $1$-dimensional and aligned with the exterior normal at the upper plate.  
Imposing symmetry along the center plane $x_\perp= 0$ and introducing the corresponding splittings $u_p = (u_\paral,u_\perp)$ and $\nabla = (\nabla_\paral,\dxperp)$, we are led to investigate the set of equations  
\begin{subequations}
	\label{propParPlates}
	\begin{align}
		\partial_t p + \delparal\cdot(cu_\paral) + \dxperp (cu_\perp) &= 0 &&\text{in }\Omega_\paral\times (0,a),\label{ppp1}\\
		\partial_t u_\paral + c\,\delparal p & = 0 &&\text{in } \Omega_\paral\times (0,a),\label{ppp2}\\
		\partial_t u_\perp + c\,\dxperp p & = 0 &&\text{in } \Omega_\paral\times (0,a),\label{ppp3}\\
		u_\perp &= 0 &&\text{on }\Omega_\paral\times \{0\},\label{ppp4}\\
		u_\perp&= - c\,\delparal\cdot\big(\sqrt{\tau_V} \fracD[-\infty]{t}{-1/2}u_{\paral}\big) +  \sqrt{\tau_T}\fracD[-\infty]{t}{1/2} p &&\text{on } \Omega_\paral\times \{a\}, \label{ppp5}\\
		p-n_\paral\cdot u_\paral &= 0 &&\text{on } \partial\Omega_\paral\times (0,a),\label{ppp6}	  		
	\end{align}
\end{subequations}
where $n_\paral$ denotes the exterior unit normal to the mantle surface $\partial\Omega_\paral\times (0,a)$.
Note that equations are posed in $t\in \mathbb{R}$, which will allow us to explicitly construct the normal modes.
Let us consider a normal mode ansatz of the form
\begin{subequations}
	\label{ansatz}
	\begin{align}
		p(t,x_\paral,x_\perp) &= q(x_\paral)\cosh(k_\perp x_\perp)\exp(st),\label{ansatzP}\\
		u_\paral(t,x_\paral,x_\perp) &=  -\frac{c}{s}\delparal p(t,x_\paral,x_\perp),\\
		u_\perp(t,x_\paral,x_\perp) &=  -\frac{c}{s}\dxperp p(t,x_\paral,x_\perp) = -\frac{ck_\perp\tanh(k_\perp x_\perp)}{s}p(t,x_\paral,x_\perp), 
	\end{align}
\end{subequations}
where $s,k_\perp\in\mathbb{C}\setminus\{0\}$.
We immediately find that ansatz~\eqref{ansatz} satisfies equations~\eqref{ppp2}--\eqref{ppp4} for each combination of the parameters, and equations~\eqref{ppp1} and~\eqref{ppp6} provided
\begin{subequations}
			\label{internalDispersion}
	\begin{align}
		- \delparal\cdot(\delparal q )&= \bigg(k_\perp^2-\frac{s^2}{c^2}\bigg)q \equiv k_\paral^2  q &&\text{in } \Omega_\paral,\label{tLapl}\\
		\frac{s}{c} \,q +  \,n_\paral\cdot \delparal q &= 0 &&\text{on } \partial\Omega_\paral,
	\end{align}
\end{subequations}
that is, $k_\paral^2 \equiv k_\perp^2-s^2/c^2$ must be an eigenvalue of the $(d-1)$-dimensional Robin Laplacian on $\Omega_\paral$ and $q$ a corresponding eigenfunction. 

Let us look for modes~\eqref{ansatz} with real positive $s$; that is, modes that are \emph{unbounded in time}.  
For positive $s$ the eigenvalue $k_\paral^2$ must be positive~\cite[§4.2]{BuFrKe17}, which implies that $k_\perp^2 = k_\paral^2 + s^2/c^2 > 0$.
Without loss of generality, we assume that $k_\paral > 0$ and, noting that the ansatz~\eqref{ansatz} is independent of the sign of $k_\perp$, also that $k_\perp> 0$.
Separation of variables with the ansatz $q(x_\paral) = \Pi_{i=1}^{d-1} X_i(x_{\paral_i})$ reduces eigenvalue problem~\eqref{internalDispersion} to a set of one dimensional eigenvalue problems,
\begin{subequations}
	\label{OneDEigen}
	\begin{align}
		-X_i''(x_{\paral_i}) &= k_i^2 X_i(x_{\paral_i}) &&0 < x_{\paral_i} < L_i,\\
		\frac{s}{c} \,X_i(0) - X_i'(0) &= 0,\\
		\frac{s}{c} \,X_i(L_i) + X_i'(L_i) &= 0,
	\end{align}
\end{subequations}
where $k_i > 0$ and $k_\paral^2 = \sum_{i = 1}^{d-1} k_i^2$.
One may verify that 
\begin{equation}
	\label{ansatzQ}
	X_i(x_{\paral_i}) \propto \cos (k_ix_{\paral_i}) + \frac{s}{ck_i}\sin (k_ix_{\paral_i})
\end{equation}
solves eigenvalue problem~\eqref{OneDEigen} provided~\cite[eq.~(4.11)]{BuFrKe17} 
\begin{equation}
	\label{sRelki1}
	s = ck_i\big[|\csc(k_iL_i)|-\cot(k_iL_i)\big].
\end{equation}
Noting that the bracketed part of expression~\eqref{sRelki1} is $\pi$-periodic and using half-angle trigonometric identities, we find a neater expression relating $s$ and $k_i$,
\begin{equation}
	\label{sRelki}
	s = ck_i\tan\left(\frac{\theta_i}{2}\right)\text{, where $k_iL_i  = \theta_i + n_i\pi$ for some $\theta_i\in[0,\pi)$ and $n_i\in\mathbb{N}$}.
\end{equation}
Substituting ansatz~\eqref{ansatz} into boundary condition~\eqref{ppp5}, recalling that $\fracD[-\infty]{t}{-1/2}\exp(st) = \exp(st)/\sqrt s$ for $s>0$~\cite[Table~9.2]{SaKiMa93} and therefore $\fracD[-\infty]{t}{1/2}\exp(st) = \partial_t \fracD[-\infty]{t}{-1/2}\exp(st) =  \exp(st)\sqrt s$, and equation~\eqref{tLapl}, we find
\begin{equation}
	-\frac{ck_\perp\tanh(k_\perp a)}{s}p(t,x_\paral,a) = -\frac{c^2k_\paral^2\sqrt{\tau_V}}{s^{3/2}}p(t,x_\paral,a) + \sqrt{s\tau_T}\,p(t,x_\paral,a), 
\end{equation}
that is, 
\begin{align}
	\label{dispBnd}
	ck_\perp \tanh(k_\perp a) - \frac{c^2k_\paral^2\sqrt{\tau_V}}{\sqrt{s}} + s^{3/2}\sqrt{\tau_T} = 0.
\end{align}
where $ck_\perp  = \sqrt{s^2 + c^2k_\paral^2}$ and $k_\paral^2 = \sum_{i = 1}^{d-1} k_i^2$.
Eliminating $s$ in equation~\eqref{dispBnd} using expressions~\eqref{sRelki}, we finally obtain a system of $d-1$ equations relating the $d-1$ eigenvalues $k_i$,
\begin{align}
	c\sqrt{k_i^2\tan^2\left(\frac{\theta_i}{2}\right) + \sum\limits_{j=1}^{d-1}k_j^2} \,\tanh\left(a\sqrt{k_i^2\tan^2\left(\frac{\theta_i}{2}\right) + \sum\limits_{j=1}^{d-1}k_j^2}\right) - \sqrt{\tau_V}\frac{c^2\sum\limits_{j=1}^{d-1}k_j^2}{\sqrt{ck_i\tan\left(\frac{\theta_i}{2}\right)}}\nonumber\\
	+ \sqrt{\tau_T}\left(ck_i\tan\left(\frac{\theta_i}{2}\right)\right)^{3/2} = \enspace 0\text{, where $k_iL_i  = \theta_i + n_i\pi$ for some $\theta_i\in(0,\pi)$ and $n_i\in\mathbb{N}$}.\label{syst}
\end{align} 

The conclusion is that if there are solutions to system~\eqref{syst}, then the corresponding ansatz~\eqref{ansatz} fulfills equations~\eqref{propParPlates}.
To facilitate the analysis, we introduce the dimensionless quantities $K_i = k_iL$, $Q_i = L/L_i$, $A = a/L$, $T_V = (c/L)\tau_V$, $T_T = (c/L)\tau_T$,  where $L = \sum_{i=1}^{d-1}L_i$. 
In these dimensionless quantities, system~\eqref{syst} can be expressed as
\begin{align}
	\sqrt{K_i^2\tan^2\left(\frac{\theta_i}{2}\right) + \sum\limits_{j=1}^{d-1}K_j^2} &\,\tanh\left(A\sqrt{K_i^2\tan^2\left(\frac{\theta_i}{2}\right) + \sum\limits_{j=1}^{d-1}K_j^2}\right)\nonumber \\
	&- \sqrt{T_V}\frac{\sum\limits_{j=1}^{d-1}K_j^2}{\sqrt{K_i\tan\left(\frac{\theta_i}{2}\right)}}
	+ \sqrt{T_T}\left(K_i\tan\left(\frac{\theta_i}{2}\right)\right)^{3/2} = \enspace 0,	\label{systNonDim}
\end{align}
where $K_i = k_i L =Q_i k_i L_i = Q_i(\theta_i + n_i\pi)$ fore some $\theta_i\in(0,\pi)$ and $n_i\in\mathbb{N}$.

\begin{lem}
	\label{lem:existence}
	Assume that $T_V > 0$.
	For each combination of $n_1, \ldots, n_{d-1} \in\mathbb{N}\setminus \{0\}$, there is at least one solution $\theta_1, \ldots, \theta_{d-1} \in (0,\pi)$ to system~\eqref{systNonDim}.
\end{lem}
\begin{proof}
	Let $n_1, \ldots, n_{d-1} \in\mathbb{N}\setminus \{0\}$ be given. 
	For each $i$, the left side of expression~\eqref{systNonDim} defines a continuous function 
	\begin{align}
		R_i: (\theta_1,\ldots, \theta_{i-1},\theta_i,\theta_{i+1},\ldots, \theta_{d-1})\in [0,\pi]^{i-1}\times (0,\pi) \times [0,\pi]^{d-1-i}\mapsto R_i(\theta_1,\ldots,\theta_{d-1})\in\mathbb{R}
	\end{align}
    with limits $\lim_{\theta_i\to 0} R_i(\theta_1,\ldots,\theta_{d-1}) = -\infty$ and $\lim_{\theta_i\to \pi} R_i(\theta_1,\ldots,\theta_{d-1}) = \infty$.
	Based on continuity, we define bounded functions $r_i: [0,\pi]^{d-1}\to [-1,1]$ by
		\begin{align}
			r_i(\theta_1,\ldots, \theta_{d-1}) = 
			\begin{cases}
				\tanh (R_i(\theta_1,\ldots, \theta_{d-1})) & \text{$\theta_i\in (0,\pi)$ and $\theta_j\in [0,\pi]$ for $j\neq i$,}\\
				-1 & \text{$\theta_i = 0$ and $\theta_j\in [0,\pi]$ for $j\neq i$,}\\
				\phantom{-}1 & \text{$\theta_i = \pi$ and $\theta_j\in [0,\pi]$ for $j\neq i$.}
			\end{cases}
		\end{align}
	Functions $r_1,\ldots, r_{d-1}$ satisfy the assumptions of the Poincaré--Miranda theorem~\cite{Ku97} and therefore there exist $\theta_1^\ast,\ldots,\theta_{d-1}^\ast \in (0,\pi)$ such that $r_1(\theta_1^\ast,\ldots,\theta_{d-1}^\ast) =\ldots = r_{d-1}(\theta_1^\ast,\ldots,\theta_{d-1}^\ast) = 0$; that is, $R_1(\theta_1^\ast,\ldots,\theta_{d-1}^\ast) =\ldots = R_{d-1}(\theta_1^\ast,\ldots,\theta_{d-1}^\ast) = 0$.
\end{proof}

Having demonstrated existence of solutions to system~\eqref{systNonDim}, we continue, assuming that $T_V, T_T > 0$, with investigating the asymptotic behavior for $n_1  = \ldots = n_{d-1} \equiv n \gg 1$; that is, the behavior for large eigenvalues $k_i$ of eigenvalue problems~\eqref{OneDEigen}. 
Retaining the most significant terms in system~\eqref{systNonDim}, recalling that $\lim_{x\to \infty}\tanh(x) = 1$ and $K_i = Q_i (\theta_i + n\pi)\approx Q_i \,n\pi$, we are led to study the system 
\begin{align}
	n\pi\sqrt{Q_i^2\tan^2\left(\frac{\theta_i}{2}\right) + \sum\limits_{j=1}^{d-1}Q_j^2} - \sqrt{T_V}\frac{\sum\limits_{j=1}^{d-1}Q_j^2}{\sqrt{Q_i\tan\left(\frac{\theta_i}{2}\right)}}(n\pi)^{3/2}
	+ \sqrt{T_T}\left(n\pi\,Q_i\tan\left(\frac{\theta_i}{2}\right)\right)^{3/2} = \enspace 0.	\label{systNonDimAsympt}
\end{align}
Multiplying with $(n\pi)^{-3/2}$ and introducing $z_i = Q_i\tan\left(\frac {\theta_i}{2}\right)>0$ and $Q = \sqrt{\sum_{j=1}^{d-1}Q_j^2} > 0$ in expression~\eqref{systNonDimAsympt}, we obtain the decoupled equations
\begin{equation}
	\label{eq_z}
		\frac{1}{\sqrt{n\pi}} \sqrt{z_i^2 + Q^2} - \sqrt{T_V}\,\frac{Q^2 }{\sqrt{z_i}}
		+ \sqrt{T_T}\,z_i^{3/2} = 0.
\end{equation}
Since the first term is positive, it is necessary that 
\begin{align}
	\sqrt{T_V}\,\frac{Q^2 }{\sqrt{z_i}} \geq \sqrt{T_T}\,z_i^{3/2} \Leftrightarrow z_i \leq Q \left(\frac{T_V}{T_T}\right)^{1/4}<\infty,
\end{align}
and therefore passing to the limit in equation~\eqref{eq_z}, we find
\begin{equation}
	\label{lim_z}
	\lim\limits_{n\to\infty} z_i = Q \left(\frac{T_V}{T_T}\right)^{1/4} = L\left(\frac{\tau_V}{\tau_T}\right)^{1/4}\sqrt{\sum_{i = 1}^{d-1}\frac{1}{L_i^2}},
\end{equation}
where we in the last step used that $T_V/T_T = \tau_V/\tau_T$, $Q = \sqrt{\sum_{i=1}^{d-1}Q_i^2}$ and $Q_i = L/L_i$.
Recalling expressions~\eqref{sRelki},  $z_i = Q_i \tan\left(\frac {\theta_i}{2}\right)$, $Q_i = L/L_i$ and $k_i L_i = n\pi + \theta_i$, we find that
\begin{align}
	\label{extraStep}
	s = c k_i \tan\left(\frac {\theta_i}{2}\right) = c k_i\frac{z_i}{Q_i} = ck_iL_i\frac{z_i}{L} = c(n\pi + \theta_i)  \frac{z_i}{L}.
\end{align}
Passing to the limit in expression~\eqref{extraStep}, recalling limit~\eqref{lim_z}, we find that asymptotically, as $n\to\infty$,
\begin{equation}
	s = c \,n\pi \left(\frac{\tau_V}{\tau_T}\right)^{1/4}\sqrt{\sum_{i = 1}^{d-1}\frac{1}{L_i^2}}.
\end{equation}
Thus, there are arbitrarily large $s>0$ for which the ansatz~\eqref{ansatz} satisfies equations~\eqref{propParPlates}.

Theorem~\ref{th:1} summarizes the findings so far in this section. 
\begin{thm}
	\label{th:1}
	Equations~\eqref{propParPlates}, with $\tau_V, \tau_T > 0$,  admit solutions that grow arbitrarily fast in time. 
\end{thm}

Having investigated equations~\eqref{propParPlates}, which model lossy sound propagation between parallel plates for $t\in\mathbb{R}$, we briefly comment on the homogeneous initial--boundary-value problem formed by supplementing these equations with homogeneous initial conditions for $p$ and $u$ at $t = 0$. 
Assuming, as in the previous section, that $p = 0$ and $u = 0$ also for $t < 0$, and formally applying the (one-sided) temporal Laplace-transform to equations~\eqref{propParPlates}, we find
\begin{subequations}
	\label{propParPlates_s}
	\begin{align}
		s \check{p} + \delparal\cdot(c\check{u}_\paral) + \dxperp (c\check{u}_\perp) &= 0 &&\text{in }\Omega_\paral\times (0,a),\label{ppps1}\\
		s \check{u}_\paral + c\,\delparal \check{p} & = 0 &&\text{in } \Omega_\paral\times (0,a),\label{ppps2}\\
		s \check{u}_\perp + c\,\dxperp \check{p} & = 0 &&\text{in } \Omega_\paral\times (0,a),\label{ppps3}\\
		\check{u}_\perp &= 0 &&\text{on }\Omega_\paral\times \{0\},\label{ppps4}\\
		\check{u}_\perp&= - c\,\delparal\cdot\left(\sqrt{\frac{\tau_V}{s}}\check{u}_{\paral}\right) +  \sqrt{s\tau_T}\check{p} &&\text{on } \Omega_\paral\times \{a\}, \label{ppps5}\\
		\check{p}-n_\paral\cdot \check{u}_\paral &= 0 &&\text{on } \partial\Omega_\paral\times (0,a),\label{ppps6}	  		
	\end{align}
\end{subequations}
where $\check p(s,x_\paral,x_\perp) = \mathcal{L}\{p(t,x_\paral,x_\perp)\}(s)$, $\check u_\paral(s,x_\paral,x_\perp) = \mathcal{L}\{u_\paral(t,x_\paral,x_\perp)\}(s)$, and $\check u_\perp(s,x_\paral,x_\perp) = \mathcal{L}\{u_\perp(t,x_\paral,x_\perp)\}(s)$ denote the Laplace-transformed fields. 
We have also used that $\mathcal{L}\{\fracD{t}{-1/2}f\}(s) = \mathcal{L}\{f(t)\}(s)/\sqrt{s}$~\cite[eq.~(7.14)]{SaKiMa93} and therefore $\mathcal{L}\{\fracD{t}{1/2}f\}(s) = \mathcal{L}\{\partial_t\,\fracD{t}{-1/2}f\}(s) =  \mathcal{L}\{f(t)\}(s)\sqrt{s}$ for functions $f$ vanishing at $t = 0$. 
Based on the time-domain ansatz~\eqref{ansatz}, we may introduce the Laplace-domain ansatz 
\begin{subequations}
	\label{ansatzLapl}
	\begin{align}
		\check p(s,x_\paral,x_\perp) &= p(0,x_\paral,x_\perp),\label{ansatzLaplP}\\
		\check u_\paral(s,x_\paral,x_\perp) &= u_\paral(0,x_\paral,x_\perp),\\ 
		\check u_\perp(s,x_\paral,x_\perp) &= u_\perp(0,x_\paral,x_\perp), 
	\end{align}
\end{subequations}
which leads to identical relations between the parameters $s, k_\perp$, and $k_i$ as in the time-domain analysis above.
Thus there are arbitrarily large (real) positive values of $s$ for which ansatz~\eqref{ansatzLapl} solves boundary-value problem~\eqref{propParPlates_s} and consequently it is impossible to find an inversion contour. 
As argued by Brambley~\cite{Br09}, such behavior of the Laplace-domain boundary-value problem is indicative of ill-posedness.  

\begin{figure}
	\centering
	\includegraphics{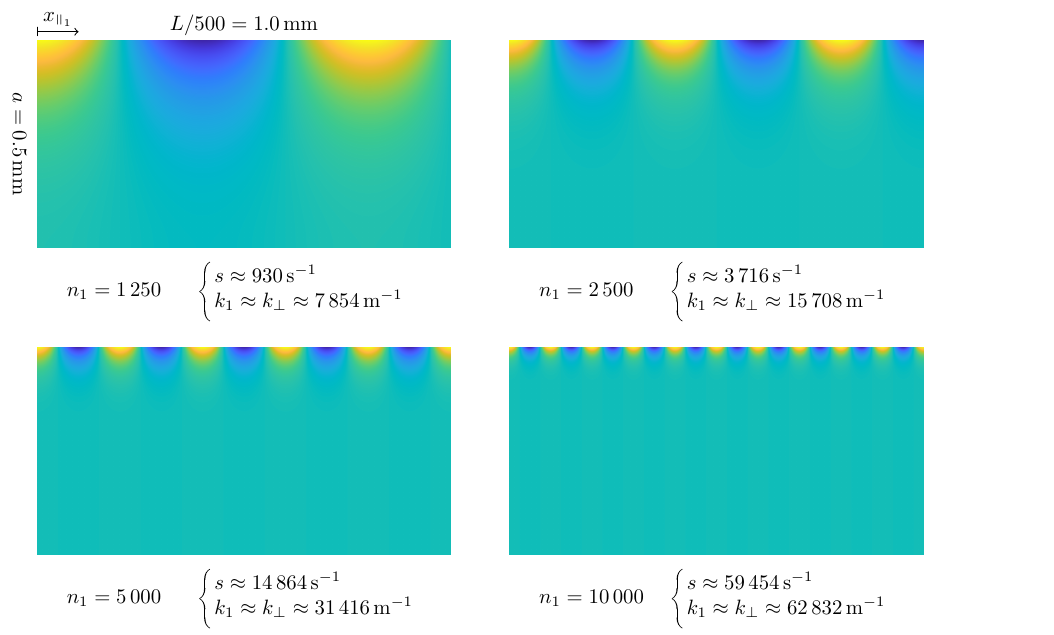}
	\caption{Snapshots at $t = 0$ of some pressure modes exhibiting exponential temporal growth in a 2D duct with dimensions $a = 0.5\,$mm and $L = 500\,$mm.}
	\label{f:modes}
\end{figure}

Figure~\ref{f:modes} presents snapshots at $t = 0$ in the vicinity of $x_{\paral_1} = 0$ of 2D pressure modes defined by expressions~\eqref{ansatzP} and~\eqref{ansatzQ}, which coincide with the Laplace-domain pressure modes~\eqref{ansatzLaplP} by construction. 
The dimensions of the duct, $a = 0.5\,$mm and $L = 500\,$mm, have been chosen to match those in the numerical experiments below. 
Inspecting the figure and formulae, we find that large amplitudes of pressure concentrate near the solid wall on the top as the wave number increases.     
These modes look strikingly similar to \textit{surface waves}~\cite[\S~3.2.4]{RiHi21}, but have a purely exponential time dependence without oscillations.   
Since the largest observable wave number in the numerical experiments is inversely proportional to the spatial discretization step, we expect absolute instabilities that become worse with grid refinements to form close to the solid wall.

\section{Finite-difference time-domain discretization}
\label{sec:FDTD}
\begin{figure}
	\centering
	\includegraphics{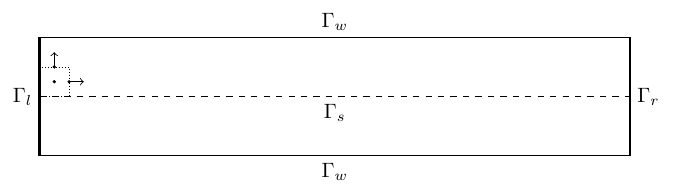}
	\caption{Finite section of a two-dimensional duct.
	The dotted square indicates the $m = 1, n=1$ cell used for discretization. The origin of the coordinate system is chosen to coincide with the lower left corner of the indicated cell.}
	\label{duct}
\end{figure}

In this section, we apply a finite-difference time-domain (FDTD) discretization of the initial--boundary-value problem~\eqref{ibp} in a finite section of a straight two-dimensional duct as illustrated in Figure~\ref{duct}, where we let $\Gamma_{io} = \Gamma_l \cup \Gamma_r$. 
Assuming that the source $g$ is symmetric about the center line $\Gamma_s$, we model only the upper part of the duct by requiring
\begin{align}
	\label{symcond}
	n\cdot u_p = 0\quad\quad \text{on }\Sigma_s = (0,T)\times \Gamma_s.
\end{align}

\begin{figure}
	\centering
	\includegraphics{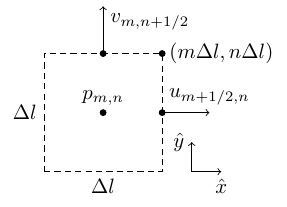}
	\caption{Illustration of the physical location of the quantities in the update scheme~\eqref{interiorUpdate}.}
	\label{Yee}
\end{figure}

Introducing the grid quantities
\begin{subequations}
	\label{gridQuantities}
	\begin{align}
		p_{m,n}^q &\sim p(q\Delta t, (m-1/2)\Delta l,(n-1/2)\Delta l),\label{pmnq}\\
		u_{m+1/2,n}^{q+1/2} &\sim \hat{x}\cdot u_p((q+1/2)\Delta t,m\Delta l,(n-1/2)\Delta l),\label{umnq}\\
		v_{m,n+1/2}^{q+1/2} &\sim \hat{y}\cdot u_p((q+1/2)\Delta t,(m-1/2)\Delta l,n\Delta l)\label{vmnq},
	\end{align}
\end{subequations}
as illustrated in Figure~\ref{Yee}, we update the pressure and velocities in the interior of the duct, $m\in\{1, \ldots, M\}, n\in \{1, \ldots, N\}$, according to the classical Yee scheme~\cite{Taflove2005cet},
\begin{subequations}
	\label{interiorUpdate}
	\begin{align}
		p_{m,n}^q &= p_{m,n}^{q-1} - \Co\left(u_{m+1/2,n}^{q-1/2}-u_{m-1/2,n}^{q-1/2} + v_{m,n+1/2}^{q-1/2}-v_{m,n-1/2}^{q-1/2}\right),\label{pUpdInt}\\
		u_{m+1/2,n}^{q+1/2} &= u_{m+1/2,n}^{q-1/2} - \Co\left(p_{m+1,n}^{q}-p_{m,n}^{q}\right),\label{uUpdInt}\\
		v_{m,n+1/2}^{q+1/2} &= v_{m,n+1/2}^{q-1/2} - \Co\left(p_{m,n+1}^{q}-p_{m,n}^{q}\right)\label{vUpdInt},
	\end{align}
\end{subequations}
where $\Co = c\Delta t/\Delta l$ denotes the Courant number.
Note that the scheme is staggered in both space and time. 

Examining update equations~\eqref{interiorUpdate}, we find that the quantities $u_{1/2,n}^{q+1/2}$, $v_{m,1/2}^{q+1/2}$, $p_{M+1,n}^{q}$, and $p_{m,N+1}^{q}$ need to be updated based on the boundary conditions. 
The symmetry condition~\eqref{symcond} on $\Gamma_s $ implies that 
\begin{equation}
	\label{vUpdLower}
	v_{m,1/2}^{q+1/2} = 0, \quad\quad m\in\{1,\ldots M\}.
\end{equation}
Differentiating boundary condition~\eqref{ibp5} with respect to time, and eliminating $\partial_t u_p$ using equation~\eqref{ibp2}, we obtain
\begin{equation}
	\label{iop}
	\partial_t p + n\cdot c\nabla p = 2\,\partial_t g \quad\quad \text{on }\Gamma_{io} = \Gamma_l \cup \Gamma_r.
\end{equation}	
The finite difference approximations
\begin{align}
	\label{fd1}
	\partial_t p &\sim \frac{\frac{1}{2}(p_{0,n}^{q}+p_{1,n}^{q}\big)-\frac{1}{2}\big(p_{0,n}^{q-1}+p_{1,n}^{q-1}\big)}{\Delta t}, \\
	n\cdot c\nabla p &\sim - c\frac{\frac{1}{2}(p_{1,n}^{q-1}+p_{1,n}^{q}\big)-\frac{1}{2}\big(p_{0,n}^{q-1}+p_{0,n}^{q}\big)}{\Delta l}
\end{align}
at $\Gamma_l$ and 
\begin{align}
	\partial_t p &\sim \frac{\frac{1}{2}(p_{M,n}^{q}+p_{M+1,n}^{q}\big)-\frac{1}{2}\big(p_{M,n}^{q-1}+p_{M+1,n}^{q-1}\big)}{\Delta t},\\ 
	n\cdot c\nabla p &\sim  c\frac{\frac{1}{2}(p_{M+1,n}^{q-1}+p_{M+1,n}^{q}\big)-\frac{1}{2}\big(p_{M,n}^{q-1}+p_{M,n}^{q}\big)}{\Delta l}
	\label{fd4}
\end{align}
at $\Gamma_r$ substituted into boundary condition~\eqref{iop} leads to updates analogous to those proposed by Mur~\cite[eq.~(15)]{Mu81} for computational electromagnetics,
\begin{align}
	p_{0,n}^{q} & = p_{1,n}^{q-1}+ \frac{1-\Co}{1+\Co}\left(p_{0,n}^{q-1}-p_{1,n}^{q}\right)+\frac{4\Delta t}{1+\Co}(\partial_t g\rvert_{\Gamma_l})_{n}^{q-1/2}, && n\in \{1,\ldots,N\},\label{pUpdLeft}\\
	u_{1/2,n}^{q+1/2} &=  u_{1/2,n}^{q-1/2} - \Co\left(p_{1,n}^{q}-p_{0,n}^{q}\right),&& n\in \{1,\ldots,N\},\label{uUpdLeft}\\
	p_{M+1,n}^{q} & = p_{M,n}^{q-1}+ \frac{1-\Co}{1+\Co}\left(p_{M+1,n}^{q-1}-p_{M,n}^{q}\right)+\frac{4\Delta t}{1+\Co}(\partial_t g\rvert_{\Gamma_r})_{n}^{q-1/2}, && n\in \{1,\ldots,N\}\label{pUpdRight},
\end{align}
where $(\partial_t g\rvert_{\Gamma_l})_{n}^{q-1/2}$  and $(\partial_t g\rvert_{\Gamma_r})_{n}^{q-1/2}$ correspond to evaluations at $t = (q-1/2)\Delta t$, $y = (n-1/2)\Delta l$.

It remains to devise an update for $p_{m,N+1}^{q}$ based on boundary condition~\eqref{ibp4}.
Applying $\fracD{t}{1/2}$ to boundary condition~\eqref{ibp4}, we obtain
\begin{equation}
	\label{pre_wp}
	n\cdot \fracD{t}{1/2}u_p  =  -c\,\nabla_\Gamma\cdot\big(\sqrt{\tau_V}\, \fracD{t}{1/2}\fracD{t}{-1/2}u_{p,\Gamma}\big) +  \sqrt{\tau_T}\,\fracD{t}{1/2}\fracD{t}{1/2}  p \quad\quad \text{on }\Gamma_{w}.
\end{equation}	
Since $p$ vanishes at $t = 0$ it holds that $ \fracD{t}{-1}\partial_t p \equiv \int_{0}^{t} \partial_\tau p(\tau)\,\mathrm{d}\tau  = p(t)$.
Thus, the semigroup property of fractional integration~\cite[Th.~2.5]{SaKiMa93} implies 
\begin{equation}
	\label{RLisC}
	\fracD{t}{1/2}p \equiv \partial_t \,\fracD{t}{-1/2} p =   \partial_t \,\fracD{t}{-1/2} \fracD{t}{-1}\partial_t p = \partial_t \,\fracD{t}{-1}\fracD{t}{-1/2}\partial_t p = \fracD{t}{-1/2}\partial_t p.
\end{equation}
That is, the Riemann--Liouville derivative coincides with the Caputo derivative in this case. 
Identity~\eqref{RLisC} and the semigroup property of fractional integration imply 
\begin{equation}
	\label{TwoHalvesIsWhole}
	\fracD{t}{1/2}\fracD{t}{1/2}p \equiv \partial_t\, \fracD{t}{-1/2}\,  \partial_t\, \fracD{t}{-1/2}p = \partial_t\, \fracD{t}{-1/2}   \fracD{t}{-1/2}\partial_tp = \partial_t \,\fracD{t}{-1} \partial_t p =  \partial_t p,
\end{equation}
which will be used to rewrite the second term to the right in boundary condition~\eqref{pre_wp}.
For the first term on the right, we proceed as follows. 
In general, it holds that~\cite[Th.~2.4]{SaKiMa93}
\begin{equation}
	\fracD{t}{1/2}\fracD{t}{-1/2}u_{p,\Gamma} =  u_{p,\Gamma},
\end{equation}
and therefore using that $u_{p,\Gamma} = \fracD{t}{-1}\partial_t u_{p,\Gamma}$ since $u_{p,\Gamma}$ vanishes at $t = 0$, we have for the first term on the right of boundary condition~\eqref{pre_wp} the identity
\begin{equation}
	\label{uGamma2p}
	\fracD{t}{1/2}\fracD{t}{-1/2}u_{p,\Gamma} =  u_{p,\Gamma} =  \fracD{t}{-1}\partial_t u_{p,\Gamma} = -\fracD{t}{-1} c\nabla_\Gamma p, 
\end{equation}
where equation~\eqref{ibp2} has been used to eliminate $u_{p,\Gamma}$ in the last step.
We now proceed with the term on the left of boundary condition~\eqref{pre_wp}. 
Since $u_p$ vanishes at $t = 0$, as already demonstrated for $p$ in expression~\eqref{RLisC}, the half order Riemann--Liouville and Caputo derivatives coincide and therefore
\begin{equation}
	\label{id3}
	\fracD{t}{1/2} u_p = \fracD{t}{-1/2}\partial_t u_p = -\fracD{t}{-1/2}\,c\nabla p,
\end{equation}
where equation~\eqref{ibp2} has been used, once more, to eliminate $u_{p,\Gamma}$ in the last step.
Thus, moving the term on the left side to the right side of boundary condition~\eqref{pre_wp}, dividing by  $\sqrt{\tau_T}$ and invoking identities~\eqref{TwoHalvesIsWhole},~\eqref{uGamma2p} and~\eqref{id3}, we finally obtain 
\begin{equation}
	\label{wp}
	\partial_t p + \frac{1}{\sqrt{\tau_T}}\fracD{t}{-1/2}(n\cdot c\nabla p)  + \sqrt{\frac{\tau_V}{\tau_T}} \fracD{t}{-1} c^2\Delta_\Gamma p = 0 \quad\quad \text{on }\Gamma_{w}.
\end{equation}	

We  note that for  $r \in C([0,T])$ 
\begin{align}
	\left|\frac{1}{\sqrt{\tau_T}}\fracD{t}{-1/2}r\right| &\leq \left(\frac{1}{\sqrt{\pi\tau_T}}\int\limits_{0}^{t}\frac{\mathrm{d}\tau}{\sqrt{t-\tau}}\right) \max\limits_{t\in[0,T]}|r(t)| = 2\sqrt{\frac{t}{\pi\tau_T}} \max\limits_{t\in[0,T]} |r(t)|, \\
	\left|\fracD{t}{-1}r\right| &\leq t \max\limits_{t\in[0,T]}|r(t)|, 
\end{align}
which indicates that boundary condition~\eqref{wp} is potentially stiff due to its second term; recall that $\tau_T$ is typically much smaller than the time period of audible sound.
Because of the potential stiffness, we propose an implicit discretization of boundary condition~\eqref{wp},
\begin{equation}
	\label{wpDiscr}
	p_{m,N+1}^{q}-p_{m,N+1}^{q-1} + \frac{\Co}{\sqrt{\tau_T}}\fracD{q\Delta t}{-1/2}(p_{m,N+1}-p_{m,N})  + \frac{\Co^2}{\Delta t}\sqrt{\frac{\tau_V}{\tau_T}} \fracD{q\Delta t}{-1} \Delta_m p_{N+1}= 0,
\end{equation}
where $\tau_T^{-1/2}\fracD{q\Delta t}{-1/2}(p_{m,N+1}-p_{m,N})$ and $(\Delta t)^{-1}\fracD{q\Delta t}{-1} \Delta_m p_{N+1}$ are to be defined below. 
Although discretization~\eqref{wpDiscr} is implicit, the resulting linear system is tridiagonal and therefore efficiently solvable.
Moreover, in the special case that $\tau_V = 0$ the system is even diagonal, which leads to the explicit update formula~\eqref{pUpdUpper}. 

We start by describing the discretization of the thermal contribution involving the half integral. 
Since the kernel is rather slowly decaying, naive discretizations of the convolution~\eqref{defID} eventually lead to prohibitively large computational cost.
The discretization proposed here is inspired by the works of Haddar et al.~\cite{HaLiMa10} and Monteghetti et al.~\cite{MoMaPi20} and relies on a so-called diffusive representation~\cite{HaLiMa10} of the half integral,
\begin{equation}
	\label{diffureprIHalf}
	\fracD{t}{-1/2}r = \frac{1}{\pi}\int\limits_{0}^\infty \frac{\varphi(t,\xi)}{\sqrt{\xi}}\,\mathrm{d}\xi,
\end{equation}
where the auxiliary state $\varphi$ is defined as
\begin{equation}
	\label{varphi}
	\varphi(t,\xi) = \int\limits_0^t \exp(-\xi(t-\tau))r(\tau)\,\mathrm{d}\tau.
\end{equation}
Starting with definition~\eqref{varphi}, we devise an update formula for the auxiliary state,
\begin{align}
	 	\varphi(t,\xi) &= \int\limits_0^t \exp(-\xi(t-\tau))r(\tau)\,\mathrm{d}\tau\nonumber\\ 
	 	&= \int\limits_0^{t-\Delta t} \exp(-\xi(t-\tau))r(\tau)\,\mathrm{d}\tau + \int\limits_{t-\Delta t}^{t} \exp(-\xi(t-\tau))r(\tau)\,\mathrm{d}\tau \nonumber\\
	 	&\approx \exp(-\xi\Delta t)\int\limits_0^{t-\Delta t} \exp(-\xi(t-\Delta t-\tau))r(\tau)\,\mathrm{d}\tau\nonumber\\
	 	 &\qquad\qquad\qquad+\frac{1}{2}\big(r(t-\Delta t)+r(t)\big)\int\limits_{t-\Delta t}^{t} \exp(-\xi(t-\tau))\,\mathrm{d}\tau \nonumber\\
	 	&=\exp(-\xi\Delta t)\varphi(t-\Delta t,\xi)+\frac{1}{2}\big(r(t-\Delta t)+r(t)\big)W(\xi\Delta t)\Delta t, \label{updvarphi}
\end{align}
where
\begin{equation}
	\label{defW}
	W(x) = \frac{1-\exp(-x)}{x},
\end{equation}
which is less than unity for $ x > 0$.
Replacing $\varphi$ with update~\eqref{updvarphi} in the expression~\eqref{diffureprIHalf}, we obtain
\begin{align}
	\frac{1}{\sqrt{\tau_T}}\fracD{t}{-1/2}r &= \frac{1}{\pi \sqrt{\tau_T}}\int\limits_0^\infty \frac{\varphi(t,\xi)}{\sqrt{\xi}}\,\mathrm{d}\xi \nonumber\\
	&\approx \frac{1}{\pi \sqrt{\tau_T}}\int\limits_0^\infty \frac{\exp(-\xi\Delta t)}{\sqrt{\xi}}\varphi(t-\Delta t,\xi)\,\mathrm{d}\xi + \big(r(t-\Delta t)+r(t)\big)\sqrt{\frac{\Delta t}{\pi\tau_T}},\label{updHalfIr}
\end{align}
since definition~\eqref{defW} implies that
\begin{align}
	\frac{1}{\sqrt{\tau_T}}\int\limits_0^\infty \frac{W(\xi\Delta t)\Delta t}{\pi\sqrt{\xi}}\,\mathrm{d}\xi &= \sqrt{\frac{\Delta t}{\tau_T}}\int\limits_0^\infty \frac{W(\xi\Delta t)}{\pi\sqrt{\xi\Delta t}}\,\mathrm{d}(\xi\Delta t) = \sqrt{\frac{\Delta t}{\tau_T}}\int\limits_0^\infty \frac{1-e^{-x}}{\pi x^{3/2}}\,\mathrm{d}x\\ 
	&= \sqrt{\frac{\Delta t}{\tau_T}}\int\limits_0^\infty \frac{d}{dx}\frac{2}{\pi}\bigg(\sqrt{\pi}\,\mathrm{erf}(\sqrt{x})-\frac{1-\exp(-x)}{\sqrt{x}}\bigg)\,\mathrm{d}x\ = 2\sqrt{\frac{\Delta t}{\pi\tau_T}} .
\end{align}

To obtain a fully discretized update formula, we follow the suggestion of Monteghetti et al.~\cite{MoMaPi20} to transform $\int_0^\infty\mathrm{d}\xi$ to $\int_{-1}^1\mathrm{d}\eta$ and employ a standard Gauss--Legendre quadrature rule of order $L$ with nodes  $\{\eta_l\}_{l = 1}^L$ and weights $\{w_{\eta_l}\}_{l = 1}^L$.
More precisely, employing the substitution $\xi\Delta t = (1+\eta)^2/(1-\eta)^{2}$, we find
\begin{align}
	 \label{GLapprox}
	 \frac{1}{\pi \sqrt{\tau_T}}\int\limits_0^\infty \frac{\exp(-\xi\Delta t)}{\sqrt{\xi}}\varphi(t-\Delta t,\xi)\,\mathrm{d}\xi &\approx\frac{1}{\pi\sqrt{\tau_T}}\sum\limits_{l = 1}^L \frac{\exp(-\xi_l\Delta t)}{\sqrt{\xi_l}}\varphi(t-\Delta t,\xi_l)w_l,
\end{align}
where $\xi_l\Delta t = (1+\eta_l)^2/(1-\eta_l)^{2}$, and $w_l\Delta t = 4(1+\eta_l)/(1-\eta_l)^{3}w_{\eta_l}$.
Note that the number of quadrature points $L$ needs to be significantly less than the number of time-steps in order for the diffusive representation to pay off. 

Based on letting $r(q\Delta t) = p_{m,N+1}^{q}-p_{m,N}^{q}$ in expressions~\eqref{updvarphi} and \eqref{updHalfIr} and employing the approximation~\eqref{GLapprox}, we suggest the following discretization of the third term in equation~\eqref{wpDiscr}, for $m\in\{1, \ldots, M\}$ and $l\in \{1, \ldots,L\}$,
\begin{align}
	\!\!\!\!\frac{1}{\sqrt{\tau_T}}\fracD{q\Delta t}{-1/2}(p_{m,N+1}-p_{m,N}) &= \frac{1}{\pi\sqrt{\tau_T}}\sum\limits_{l = 1}^L \frac{\exp(-\xi_l\Delta t)}{\sqrt{\xi_l}}\varphi_{m,l}^{q-1}w_l\nonumber\\ 
	&\quad+ \big(p_{m,N+1}^{q-1}-p_{m,N}^{q-1} + p_{m,N+1}^{q}-p_{m,N}^{q}\big)\sqrt{\frac{\Delta t}{\pi\tau_T}},\label{updHalfIrDiscr}\\
	\varphi_{m,l}^q &= \exp(-\xi_l\Delta t)\varphi_{m,l}^{q-1}\nonumber\\ &\quad+ \frac{1}{2}\big(p_{m,N+1}^{q-1}-p_{m,N}^{q-1} + p_{m,N+1}^{q}-p_{m,N}^{q}\big)W(\xi_l\Delta t)\Delta t, \label{updvarphiDiscr}
\end{align}
where $\varphi_{m,l}^q \approx \varphi(q\Delta t,\xi_l)$. 

We proceed with detailing the discretization of the viscous part of boundary condition~\eqref{wpDiscr}. 
Guided, once more, by the frequency-domain well-posedness theory~\cite{BeBeNo18}, we impose a homogeneous Neumann condition at $\partial \Gamma_{w}$, as displayed in equation~\eqref{ass1}, and employ a standard second order centered difference approximation in space of the tangential Laplacian,
\begin{equation}
	\Delta_\Gamma p \sim \frac{1}{\Delta l^2}\Delta_m p_{N+1}^q \equiv 
	\begin{dcases}
	\dfrac{-p_{1,N+1}^q+p_{2,N+1}^q}{\Delta l^2}& m = 1,\\
	\dfrac{p_{m-1,N+1}^q-2p_{m,N+1}^q+p_{m+1,N+1}^q}{\Delta l^2} & m\in\{2,\ldots,M-1\},\\
	\dfrac{p_{M-1,N+1}^q-p_{M,N+1}^q}{\Delta l^2} & m = M.
	\end{dcases}
\end{equation}
Combining the spatial discretization with an update strategy based on the splitting $\fracD{t}{-1} = \int_{0}^{t}\mathrm{d}\tau = \int_{0}^{t-\Delta t}\mathrm{d}\tau + \int_{t-\Delta t}^{t}\mathrm{d}\tau$ and the trapezoidal rule, we propose the following implementation of the fourth term in equation~\eqref{wpDiscr}, 
\begin{align}
	\label{updVisc}
	\frac{1}{\Delta t}\fracD{q\Delta t}{-1} \Delta_m p_{N+1} \equiv \psi_m^{q} = \psi_m^{q-1} + \frac{1}{2}\big(\Delta_m p_{N+1}^{q-1}+\Delta_m p_{N+1}^{q}\big).
\end{align}

Introducing expressions~\eqref{updHalfIrDiscr} and~\eqref{updVisc} into the discretized boundary condition~\eqref{wpDiscr} leads to a tridiagonal linear system of equations that can be numerically solved for $p_{m,N+1}^q$.
In the particular case $\tau_V = 0$, we find the explicit update formula
\begin{align}
	p_{m,N+1}^{q} &= \frac{1-\Co\sqrt{\frac{\Delta t}{\pi\tau_T}}}{1+\Co\sqrt{\frac{\Delta t}{\pi\tau_T}}}p_{m,N+1}^{q-1} + \frac{\Co\sqrt{\frac{\Delta t}{\pi\tau_T}}}{1+\Co\sqrt{\frac{\Delta t}{\pi\tau_T}}}\big(p_{m,N}^{q-1} + p_{m,N}^{q} \big)\nonumber\\ 
	&- \frac{\Co}{1+\Co\sqrt{\frac{\Delta t}{\pi\tau_T}}}\frac{1}{\sqrt{\tau_T}}\sum\limits_{l = 1}^L \frac{\exp(-\xi_l\Delta t)}{\sqrt{\xi_l}}\varphi_{m,l}^{q-1}w_l,\quad\quad m\in\{1,\ldots, M\}.\label{pUpdUpper}
\end{align}

The updates presented above, lead to the following FDTD scheme for solving initial--boundary-value problem~\eqref{ibp}:
\begin{enumerate}
	\item Initialize all variables to $0$.
	\item Update the pressure
	\begin{enumerate}
		\item at internal nodes according to expression~\eqref{pUpdInt}, and
		\item at exterior nodes adjacent to the left and right boundaries according to expressions~\eqref{pUpdLeft} and~\eqref{pUpdRight}, and
		\item at exterior nodes adjacent to the upper boundary by solving equation~\eqref{wpDiscr}, or according to expression~\eqref{pUpdUpper} if $\tau_V = 0$.
	\end{enumerate}
	\item Update the auxiliary states ($\phi$ and $\psi$) according to expressions~\eqref{updvarphiDiscr} and~\eqref{updVisc}, respectively.
	\item Update the horizontal velocity
	\begin{enumerate}
		\item at internal nodes according to expression~\eqref{uUpdInt}, and
		\item at exterior nodes adjacent to the  left boundary according to expression~\eqref{uUpdLeft}.
	\end{enumerate} 
	\item Update the vertical velocity
	\begin{enumerate}
		\item at internal nodes according to expression~\eqref{vUpdInt}, and
		\item at exterior nodes adjacent to the lower boundary according to expression~\eqref{vUpdLower}.
	\end{enumerate}
	\item Repeat from step 2. 
\end{enumerate}

\section{Numerical experiments}
\label{sec:NumExp}
\begin{figure}
	\centering
	\includegraphics[width = 17cm]{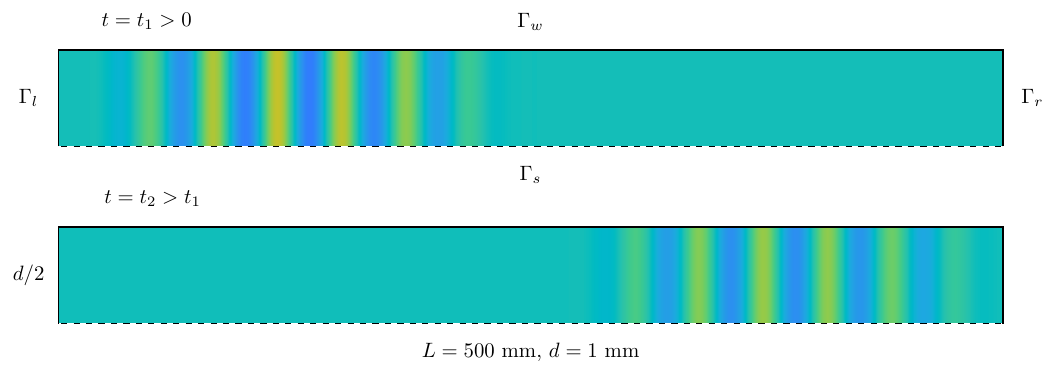}
	\caption{Two snapshots of a wave packet traveling from left to right in a straight duct with thermal boundary losses. 
		The duct has been simulated in planar symmetry; the upper boundary is a solid wall and the lower  a symmetry line.}\label{f:prop}
\end{figure}

We start with presenting numerical results obtained using the FDTD scheme for the particular case $\tau_V = 0$; that is, propagation of sound in a duct subject to thermal boundary losses only.
Based on the theoretical investigation in Section~\ref{sec:TDVTBC}, we anticipate the numerical simulations to be well-behaved in this case. 
Figure~\ref{f:prop} displays two snapshots of a wave packet traveling from left to right in an air-filled duct.
Relevant material properties are found in Table~\ref{t:matProp}, Appendix~\ref{A:matProp}.
Gauss--Legendre quadrature nodes and weights are computed using the implementation by von~Winckel~\cite{lgw}. 
The wave packet is excited using the source  
\begin{equation}
	\label{g}
	g(t) = 
	\begin{cases}
		g_l(t) = p_0\chi\left(\frac{2t-\tau_0}{\tau_0}\right)\cos(2\pi f_0 t)& \text{on }\Gamma_l,\\
		g_r(t)=0 &\text{on }\Gamma_r,
	\end{cases}
\end{equation}
where $p_0$ denotes the amplitude, $f_0 = 10$ kHz the carrier frequency, $\tau_0 = L/(4c)\approx 0.7$ ms the duration of the signal at the left boundary, and 
\begin{equation}
	\chi(x)   = 
	\begin{cases}
		\exp\left(-\frac{x^2}{1-x^2}\right), &|x|<1,\\
		0, & |x|\geq 1,
	\end{cases} 
\end{equation}
the envelope.

Aiming for a frequency resolution of $20$ Hz, we let $T = 1/20$ s and consider a sequence of discretizations with $\Delta l = 2^{-k}\Delta l_0$ and $\Delta t = 2^{-k}\Delta t_0$, where $k\in \mathbb{N}$,
 $\Delta l_0 = d/10 = 0.1 $ mm,  and $\Delta t_0 = T/245143 \approx 0.2$ \textmu s.
For these discretizations, we have $\Co \approx 0.99/\sqrt{2}$, that is, slightly below the CFL-limit for the Yee-scheme applied to the two-dimensional wave equation without boundaries. 
As mentioned above, the update~\eqref{pUpdUpper} is computationally advantageous provided that the number of quadrature points $L$ is significantly less than the number of time steps $T/\Delta t$. 
Based on initial numerical experiments, we propose $L = [5\ln(T/\Delta t)]$, where $[\cdot]$ indicates rounding to the nearest integer.
For the coarsest discretization we have $L = 62$ compared to $T/\Delta t = 245\,143$. 
 
\begin{figure}
	\centering
	\begin{subfigure}[t]{0.48\textwidth}
		\centering
	    \includegraphics[width = \textwidth]{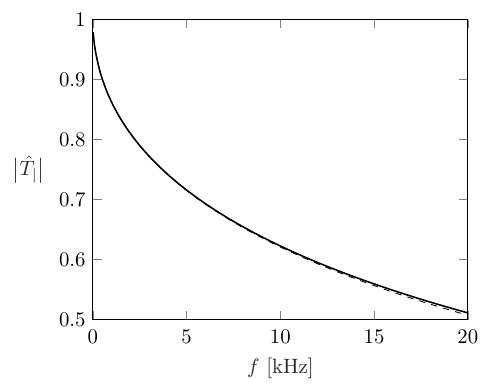}
		\caption{The solid line corresponds to the reference values; the dashed line corresponds to values based on FDTD simulations with cell width $\Delta l = \Delta l_0$.}\label{f:trans}
	\end{subfigure}
~	
	\begin{subfigure}[t]{0.48\textwidth}
		\centering
		\includegraphics[width = \textwidth]{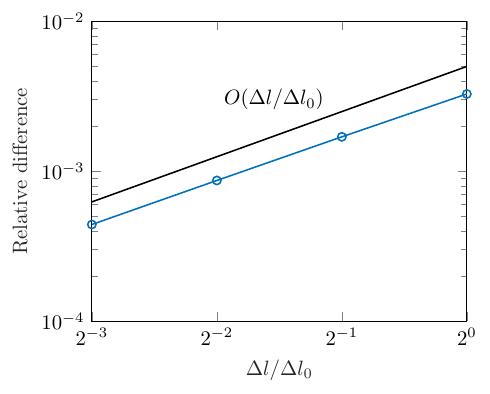}
		\caption{Relative difference to the reference in the $l_2$-norm for different cell widths~$\Delta l$.}\label{f:conv}
	\end{subfigure}
	\caption{The modulus of the transmission coefficient~\eqref{transCoeff} computed using the proposed FDTD scheme and discrete Fourier transforms compared to reference values obtained from a sequence of frequency-domain simulations in COMSOL Multiphysics employing the frequency-domain boundary condition~\eqref{harmBLBC} with $\tau_V = 0$.}
\end{figure}

To asses the proposed FDTD scheme, we have performed a sequence of reference frequency-domain simulations in COMSOL Multiphysics using boundary condition~\eqref{harmBLBC} with $\tau_V = 0$; that is, employing the computational model proposed by Berggren et al.~\cite{BeBeNo18}. 
We choose to compare the transmission characteristics for the planar mode. 
To that end, we project the outgoing characteristic at the right boundary onto the planar mode, 
\begin{align}
	\label{signal_t}
	\zeta_{\rvert}^\mathrm{t} = \frac{1}{|\Gamma_r|}\int\limits_{\Gamma_r} (p + n\cdot u_p)\,\mathrm{d}\Gamma = \frac{4}{d}\int\limits_{\Gamma_r} p\,\mathrm{d}\Gamma,
\end{align}
where the last equality follows from boundary condition~\eqref{ibp5}, since $g = g_r \equiv 0$ on $\Gamma_r$ by expression~\eqref{g}, and that $|\Gamma_r| = d/2$.
Analogously, we project the incoming characteristic at the left boundary onto the planar mode,
\begin{align}
	\label{signal_i}
	\zeta_{\rvert}^\mathrm{i} = \frac{1}{|\Gamma_l|}\int\limits_{\Gamma_l} (p - n\cdot u_p)\,\mathrm{d}\Gamma = 2g_l,
\end{align}
where the last equality follows from boundary condition~\eqref{ibp5} and expression~\eqref{g}.
We define the complex-valued transmission coefficient as the quotient of the Fourier transformed signals~\eqref{signal_t} and~\eqref{signal_i}, 
\begin{equation}
	\label{transCoeff}
	\hat T_\rvert = \frac{\hat\zeta_{\rvert}^\mathrm{t}}{\hat\zeta_{\rvert}^\mathrm{i}},
\end{equation}
which can be approximated by applying the discrete Fourier transform and numerical quadrature to grid quantities obtained from the FDTD discretization.
Figure~\ref{f:trans} demonstrates that the values of the transmission coefficients obtained using FDTD simulations on the coarse grid with cell width $\Delta l = \Delta l_0$ are in excellent agreement with the reference values obtained from a sequence of frequency-domain simulations in COMSOL Multiphysics at the discrete frequencies $f\in\{20,40,\ldots,20\,000\}$ Hz. 
As a demonstration of the convergence behavior of the FDTD simulations with respect to grid refinements, we display the relative difference of the modulus of the transmission coefficient measured in the $l_2$ norm of the values at frequencies $f\in\{20,40,\ldots,20\,000\}$ Hz in Figure~\ref{f:conv}.
Examining the relative differences more carefully reveals that convergence is deteriorating or even lacking at low frequencies. 
However, we believe that the lack of convergence at low frequencies is of little practical relevance; the frequency-domain model displays the wrong asymptotics for low frequencies anyway~\cite[Fig.~3.]{BeBeNo18}, and the maximum relative difference for frequencies below $9$ kHz is less than $10^{-4}$ for the investigated cell widths. 
Still, if needed, we expect that increasing the simulation time $T$ would improve the convergence at low frequencies at a cost, however, of increased computational demand.

\begin{figure}
	\centering
	\includegraphics{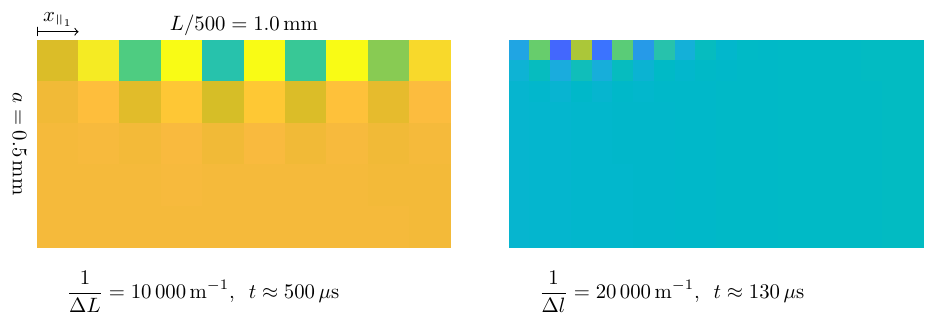}
	\caption{Snapshots of pressure capturing the formation of an absolute instability in a 2D duct with dimensions $a = 0.5\,$mm and $L = 500\,$mm.}
	\label{f:numinstab}
\end{figure}

We now proceed with the general case $\tau_V > 0$; that is, propagation of sound in a duct with both thermal and viscous boundary losses. 
The same material properties, dimensions of the duct, and source as for the case $\tau_V = 0$ are used. 
Based on the theoretical investigations in Section~\ref{sec:TDVTBC}, we expect the numerical simulations to be ill behaved in this case.  
Indeed, an absolute instability that becomes worse with mesh refinement forms at the upper boundary.
This is illustrated in Figure~\ref{f:numinstab}, which depicts snapshots of the pressure in the leftmost part of the duct. 
Note that the two snapshots are taken at different times, so their background color (pressure) differ as a result of the propagation of the wave packet.
Letting the simulations continue, we observe that the amplitude of the observed spatial grid level oscillation grows exponentially in time, and that the growth rate increases with mesh refinement.  
The close resemblance between simulations and theory leads us to the conclusion that the observed instability is most likely a property of the underlying model.   

The observant reader may have found the particular form~\eqref{wp} of boundary condition~\eqref{ibp4} somewhat revealing of the issue; applying $\partial_t$, we find that the highest order differential operators constitute a scaled space--time Laplacian, 
\begin{equation}
	\partial_t^2 + \sqrt{\frac{\tau_V}{\tau_T}}c^2 \Delta_\Gamma,
\end{equation}
which, as is well known, is \emph{not} boundedly invertible for Cauchy data, that is, for initial and boundary data. 
Thus, from a mathematical perspective and in line with the closing paragraph of Section~\ref{sec:TDpass}, it appears as if a final condition on the boundary is lacking.  
We believe that this perplexing issue and its possible resolution deserve a separate study.

\section{Acknowledgments}
Funding for this work was provided by the Swedish Research Council, grant 2018-03546. 

\newpage
\appendix

\section{Material properties of air}
\label{A:matProp}
\begin{table}[h!]
	\renewcommand{\arraystretch}{1.2}
	\caption{Properties of air at atmospheric conditions.}
	\begin{tabular}{lll}
		\hline
		mass density & $\rho_0$ & $1.204 \text{ kg} \cdot \text{m}^{-3}$\\
		kinematic viscosity & $\nu$ & $1.506\cdot 10^{-5} \text{ m}^2 \cdot \text{s}^{-1}$\\
		speed of sound & $c$ & $343.20 \text{ m} \cdot \text{s}^{-1}$\\
		specific heat at constant pressure & $c_p$ & $1.0054 \cdot 10^{3} \text{ J}\cdot \text{kg}^{-1} \cdot \text{K}^{-1}$\\
		heat capacity ratio & $\gamma$ & $1.4$\\
		\hline
		\label{t:matProp}
	\end{tabular}
\end{table}

\section{Expansions of quadratic forms involving fractional integrals}
\label{sec:expansion}

\begin{defi}
	The left Riemann--Liouville fractional integral of order $\alpha > 0$ starting at $0$ is defined by~\cite[Def.~(2.1)]{SaKiMa93}
	\begin{equation}
		\label{defIa}
		\fracD{t}{-\alpha}v = \frac{1}{\Gamma(\alpha)}\int\limits_{0}^{t} (t-\tau)^{\alpha-1}v(\tau)\,\mathrm{d}\tau.
	\end{equation}
	The right Riemann-Liouville fractional integral of order $\alpha > 0$ ending at $T$ is defined by~\cite[Def.~(2.1)]{SaKiMa93}
	\begin{equation}
		\label{defITildea}
		\fracDtilde[t]{T}{-\alpha}v = \frac{1}{\Gamma(\alpha)}\int\limits_{t}^{T} (\tau-t)^{\alpha-1}v(\tau)\,\mathrm{d}\tau.
	\end{equation}
\end{defi}

\begin{thm}
	\label{th:expansion}
	If $\alpha \in (1,\infty)\setminus \mathbb{N}$, then 
	\begin{align}
		\label{exp}
		\!\!\!\!\beta_{\alpha}\left(\frac{2}{T}\right)^{\alpha-1}\!\!\int\limits_0^T \!v(t) \fracD{t}{-\alpha}v\,\mathrm{d}t = \!\left(\int\limits_0^T v(t)\,\mathrm{d}t \right)^2\!\! - \sum\limits_{n = 1}^\infty \left(\frac{2n}{\alpha-1}-1\right)\!\!\!\left(\int\limits_0^T P_n^{\left(\frac{\alpha-1}{2}\right)}\!\!\left(\frac{2t}{T}-1\right)v(t)\,\mathrm{d}t \right)^{\!\!2}\!, 
	\end{align}
	where $P_n^{\vartheta}$ denotes the $n$th order Gegenbauer poynomial with parameter $\vartheta$ and
	\begin{equation}
		\label{Ba}
		\beta_{\alpha} = \frac{2\Gamma(\alpha)\,\Gamma\!\left(\frac{1}{2}\right)}{\Gamma\!\left(\frac{\alpha}{2}\right)\Gamma\!\left(\frac{3-\alpha}{2}\right)}.
	\end{equation}
\end{thm}
\begin{proof}
	The fractional integration by parts formula~\cite[eq.~(2.20)]{SaKiMa93},
	\begin{equation}
		\int\limits_0^T v(t)\, \fracD{t}{-\alpha}u\,\mathrm{d}t = \int\limits_0^T u(t)\,\fracDtilde[t]{T}{-\alpha}v\,\mathrm{d}t,
	\end{equation}	
	 and definitions~\eqref{defIa} and~\eqref{defITildea} imply 
	\begin{align}
		\int\limits_0^T v(t) \fracD{t}{-\alpha}v\,\mathrm{d}t &= \int\limits_0^T v(t) \left(\frac{1}{2}\,\fracD{t}{-\alpha}v+\frac{1}{2}\,\fracDtilde[t]{T}{-\alpha}v\right)\,\mathrm{d}t \nonumber\\
		&= \frac{1}{2\Gamma(\alpha)}\int\limits_0^T v(t) \left(\int\limits_{0}^{t} (t-\tau)^{\alpha-1}v(\tau)\,\mathrm{d}\tau+\int\limits_{t}^{T} (\tau-t)^{\alpha-1}v(\tau)\,\mathrm{d}\tau\right)\mathrm{d}t \nonumber\\
		& = \frac{1}{2\Gamma(\alpha)}\int\limits_0^T v(t) \left(\int\limits_{0}^{T} |t-\tau|^{\alpha-1}v(\tau)\,\mathrm{d}\tau\right)\mathrm{d}t.\label{quadForm}
	\end{align}

	Pólya and Szegö~\cite[p.~27, Hilfssatz~II]{PoSz31} have demonstrated that $|s-\zeta|^{\lambda}$, $s,\zeta\in[-1,1]$ and $\lambda>0$, has the following absolutely and uniformly convergent expansion in Gegenbauer polynomials, 
	\begin{equation}
		\label{PoSz}
		|s-\zeta|^\lambda = \frac{\Gamma\!\left(\frac{1+\lambda}{2}\right)\Gamma\!\left(1-\frac{\lambda}{2}\right)}{\Gamma\!\left(\frac{1}{2}\right)}\sum\limits_{n = 0}^\infty \left(1-\frac{2n}{\lambda}\right)P_n^{\left(-\frac{\lambda}{2}\right)}(s)P_n^{\left(-\frac{\lambda}{2}\right)}(\zeta).
	\end{equation}
	Setting $s = 2t/T-1$, $\zeta = 2\tau/T-1$ and $\lambda = \alpha-1$ in expansion~\eqref{PoSz} and recalling definition~\eqref{Ba}, we find
	\begin{align}
		\!\!\!\left(\frac{2}{T}\right)^{\alpha-1}|t-\tau|^{\alpha-1} &= \frac{\Gamma\!\left(\frac{\alpha}{2}\right)\Gamma\!\left(\frac{3-\alpha}{2}\right)}{\Gamma\!\left(\frac{1}{2}\right)}\sum\limits_{n = 0}^\infty \left(1-\frac{2n}{\alpha-1}\right)P_n^{\left(-\frac{\alpha-1}{2}\right)}\left(2t/T-1\right)P_n^{\left(-\frac{\alpha-1}{2}\right)}\left(2\tau/T-1\right)\nonumber\\
		&= \frac{2\Gamma(\alpha)}{\beta_{\alpha}}\left(1 + \sum\limits_{n = 1}^\infty \left(1-\frac{2n}{\alpha-1}\right)P_n^{\left(-\frac{\alpha-1}{2}\right)}\left(2t/T-1\right)P_n^{\left(-\frac{\alpha-1}{2}\right)}\left(2\tau/T-1\right)\right), \label{PoSz0T}
	\end{align}
	since $P_0^{(\vartheta)} \equiv 1$~\cite[§4]{Sz75}.
	Inserting expansion~\eqref{PoSz0T} in expression~\eqref{quadForm} and rearranging, we obtain the desired expansion. 
	
\end{proof}

\raggedright
\bibliographystyle{plain}
\bibliography{references}

\begin{thebibliography}{10}

\bibitem{BaBr18}
J.~S. Bach and H.~Bruus.
\newblock Theory of pressure acoustics with viscous boundary layers and streaming in curved elastic cavities.
\newblock {\em The Journal of the Acoustical Society of America}, 144(2):766--784, 2018.

\bibitem{BeWo66}
E.J. Beltrami and M.R. Wohlers.
\newblock {\em Distributions and the boundary values of analytic functions}.
\newblock Academic Press, New York, 1966.

\bibitem{BeBeNo18}
M.~Berggren, A.~Bernland, and D.~Noreland.
\newblock Acoustic boundary layers as boundary conditions.
\newblock {\em Journal of Computational Physics}, 371:633--650, 2018.

\bibitem{BiTiGaVe21}
R.~Billard, G.~Tissot, G.~Gabard, and M.~Versaevel.
\newblock Numerical simulation of perforated plate liners: Analysis of the visco--thermal dissipation mechanisms.
\newblock {\em The Journal of the Acoustical Society of America}, 149(16):16--21, 2021.

\bibitem{BoJoBr03}
R.~Bossart, N.~Joly, and M.~Bruneau.
\newblock Hybrid numerical and analytical solutions for acoustic boundary problems in thermo-viscous fluids.
\newblock {\em Journal of Sound and Vibration}, 263(1):69--84, 2003.

\bibitem{Br09}
E.~J. Brambley.
\newblock Fundamental problems with the model of uniform flow over acoustic linings.
\newblock {\em Journal of Sound and Vibration}, 322(4):1026--1037, 2009.

\bibitem{BuFrKe17}
D.~Bucur, P.~Freitas, and J.~Kennedy.
\newblock {\em 4 {T}he {R}obin problem}, pages 78--119.
\newblock De Gruyter Open Poland, Warsaw, Poland, 2017.

\bibitem{ChTh20}
J.~Chabassier and A.~Thibault.
\newblock {Viscothermal models for wind musical instruments}.
\newblock Research Report RR-9356, {Inria Bordeaux Sud-Ouest}, August 2020.

\bibitem{ChLiGr13}
L.~Cheng, Y.~Li, and K.~Grosh.
\newblock Including fluid shear viscosity in a structural acoustic finite element model using a scalar fluid representation.
\newblock {\em Journal of Computational Physics}, 247:248--261, 2013.

\bibitem{CoDaMaBaBe20}
M.~J. Cops, J.~G. McDaniel, E.~A. Magliula, D.~J. Bamford, and M.~Berggren.
\newblock Estimation of acoustic absorption in porous materials based on viscous and thermal boundary layers as acoustic boundary conditions.
\newblock {\em The Journal of the Acoustical Society of America}, 148(3):1624--1635, 2020.

\bibitem{Cr48}
L.~Cremer.
\newblock {\"U}ber die akustische {G}renzschicht vor starren {W}{\"a}nden.
\newblock {\em Archiv der Elektrischen {\"U}bertragung}, 2:136--139, 1948.

\bibitem{HaLiMa10}
H.~Haddar, J.-R. Li, and D.~Matignon.
\newblock Efficient solution of a wave equation with fractional-order dissipative terms.
\newblock {\em Journal of Computational and Applied Mathematics}, 234(6):2003--2010, 2010.
\newblock Eighth International Conference on Mathematical and Numerical Aspects of Waves (Waves 2007).

\bibitem{JiSa18}
J.~Jith and S.~Sarkar.
\newblock Boundary layer impedance model to analyse the visco-thermal acousto-elastic interactions in centrifugal compressors.
\newblock {\em Journal of Fluids and Structures}, 81:179--200, 2018.

\bibitem{Jo10}
N.~Joly.
\newblock Finite element modeling of thermoviscous acoustics on adapted anisotropic meshes: Implementation of the particle velocity and temperature variation formulation.
\newblock {\em Acta Acustica united with Acustica}, 96(1):102--114, 2010.

\bibitem{KaWiBo11}
W.~R. Kampinga, Y.~H. Wijnant, and A.~de~Boer.
\newblock An efficient finite element model for viscothermal acoustics.
\newblock {\em Acta Acustica united with Acustica}, 97(4):618--631, 2011.

\bibitem{Ki68}
G.~Kirchhoff.
\newblock Ueber den {E}influss der {W}{\"a}rmeleitung in einem {G}ase auf die {S}challbewegung.
\newblock {\em Annalen der Physik und Chemie}, 6:177--198, 1868.

\bibitem{Ku97}
W.~Kulpa.
\newblock The {P}oincar{\'e}--{M}iranda theorem.
\newblock {\em The American Mathematical Monthly}, 104(6):545--550, 1997.

\bibitem{Mo18}
F.~Monteghetti.
\newblock {\em {Analysis and Discretization of Time-Domain Impedance Boundary Conditions in Aeroacoustics}}.
\newblock Theses, {Institut Sup{\'e}rieur de l'A{\'e}ronautique et de l'Espace (ISAE-SUPAERO); Universit{\'e} de Toulouse}, October 2018.

\bibitem{MoMaPiPa16}
F.~Monteghetti, D~Matignon, E.~Piot, and L.~Pascal.
\newblock Design of broadband time-domain impedance boundary conditions using the oscillatory--diffusive representation of acoustical models.
\newblock {\em The Journal of the Acoustical Society of America}, 140(3):1663--1674, 2016.

\bibitem{MoMaPi20}
Florian Monteghetti, Denis Matignon, and Estelle Piot.
\newblock Time-local discretization of fractional and related diffusive operators using {G}aussian quadrature with applications.
\newblock {\em Applied Numerical Mathematics}, 155:73--92, 2020.
\newblock Structural Dynamical Systems: Computational Aspects held in Monopoli (Italy) on June 12-15, 2018.

\bibitem{Mu81}
G.~Mur.
\newblock Absorbing boundary conditions for the finite-difference approximation of the time-domain electromagnetic-field equations.
\newblock {\em IEEE Transactions on Electromagnetic Compatibility}, EMC-23(4):377--382, 1981.

\bibitem{Ni10}
M.~Nijhof.
\newblock {\em Viscothermal wave propagation}.
\newblock PhD thesis, University of Twente, Netherlands, December 2010.

\bibitem{Pi89}
A.~Pierce.
\newblock {\em Acoustics: An Introduction to Its Physical Principles and Applications}.
\newblock Acoustical Society of America, Melville, New York, 1989.

\bibitem{PoSz31}
G.~Pólya and G.~Szegö.
\newblock {Ü}ber den transfiniten {D}urchmesser ({K}apazitätskonstante) von ebenen und räumlichen {P}unktmengen.
\newblock {\em Journal für die reine und angewandte {M}athematik}, 165:4--49, 1931.

\bibitem{RiHaPaScBuTh11}
Ch. Richter, J.~Abdel Hay, {\l}.~Panek, N.~Sch{\"o}nwald, S.~Busse, and F.~Thiele.
\newblock A review of time-domain impedance modelling and applications.
\newblock {\em Journal of Sound and Vibration}, 330(16):3859--3873, 2011.
\newblock Computational aero-acoustics (CAA) for aircraft noise prediction~---~part A.

\bibitem{RiHi21}
S.~W. Rienstra and A.~Hirschberg.
\newblock An introduction to acoustics.
\newblock Revised and updated version of reports IWDE 92-06 and IWDE 01-03, Eindhoven University of Technology, 2021.

\bibitem{SaKiMa93}
S.~G. Samko, A.~A. Kilbas, and O.~I. Marichev.
\newblock {\em Fractional Integrals and Derivatives}.
\newblock Gordon Breach, 1993.

\bibitem{ScThJo14}
K.~Schmidt, A.~Thöns-Zueva, and P.~Joly.
\newblock High-order asymptotic expansion for the acoustics in viscous gases close to rigid walls.
\newblock {\em Mathematical Models and Methods in Applied Sciences}, 24(09):1823--1855, 2014.

\bibitem{ScTh21}
Kersten Schmidt and Anastasia Thöns-Zueva.
\newblock Impedance boundary conditions for acoustic time-harmonic wave propagation in viscous gases in two dimensions.
\newblock {\em Mathematical Methods in the Applied Sciences}, 45(12):7404--7425, 2022.

\bibitem{Sz75}
Gabor Szeg{\"o}.
\newblock Orthogonal polynomials, vol. 23.
\newblock In {\em American Mathematical Society Colloquium Publications}, 1975.

\bibitem{Taflove2005cet}
A.~Taflove and S.~C. Hagness.
\newblock {\em Computational electrodynamics: the finite-difference time-domain method}.
\newblock Artech House, Boston, 3. ed. edition, 2005.

\bibitem{lgw}
G.~von Winckel.
\newblock Legendre--{G}auss quadrature weights and nodes.
\newblock https://www.mathworks.com/matlabcentral/fileexchange/4540-legendre-gauss-quadrature-weights-and-nodes.
\newblock MATLAB Central File Exchange. Retrieved 2022-03-07.

\bibitem{Ze63}
A~Zemanian.
\newblock An {N}-port realizability theory based on the theory of distributions.
\newblock {\em IEEE Transactions on Circuit Theory}, 10(2):265--274, 1963.

\end{thebibliography}

\end{document}